\documentclass{aa}  
\usepackage{graphicx}
\usepackage{txfonts}
\usepackage[colorlinks,allcolors=blue,bookmarks=false,hypertexnames=true]{hyperref} 
\usepackage{lscape}
\titlerunning{Properties of the cores and filaments in the Ophiuchus molecular cloud}
\authorrunning{Jia et al.}
\begin{document}
   
   \title{Properties of the cores and filaments in the Ophiuchus molecular cloud and its \object{L1688} hub-filament system}

   \author{Bo-Sheng Jia\inst{1,2,9,10}
          \and
          Guo-Yin Zhang\inst{2}
          \and
          Alexander Men'shchikov\inst{3}
          \and
          Sami Dib\inst{4}
          \and 
          Jin-Zeng Li\inst{2}
          \and 
          Ke Wang\inst{5}
          \and
          Di Li\inst{6,2,7}
          \and
          Xue-Mei Li\inst{2}
          \and 
          Zhi-Yuan Ren\inst{2}
          \and
          Chang Zhang\inst{8}
          \and 
          Nageen Pervaiz\inst{2}
          \and 
          Lin Xiao\inst{1,9,10}
          }

   \institute{
         Department of Physics, Hebei University, Baoding, 071002, China
         \and
         National Astronomical Observatories, Chinese Academy of Sciences, A20 Datun Road, Chaoyang District, Beijing 100101, China\\
         \email{zgyin@nao.cas.cn;ljz@nao.cas.cn}
         \and 
         Universit\'{e} Paris-Saclay, Universit\'{e} Paris Cit\'{e}, CEA, CNRS, AIM, 91191, Gif-sur-Yvette, France\\
         \email{alexander.menshchikov@cea.fr}
         \and 
         Max Planck Institute for Astronomy, K\"{o}nigstuhl 17, 69117, Heidelberg, Germany
         \and 
         Kavli Institute for Astronomy and Astrophysics, Peking University, 5 Yiheyuan Road, Haidian District, Beijing 100871, China
         \and
         New Cornerstone Science Laboratory, Department of Astronomy, Tsinghua University, Beijing 100084, China
         \and 
         Zhejiang Lab, Hangzhou, Zhejiang 311121, China
         \and 
         School of Aerospace Science, Harbin Institute of Technology, Shenzhen 518055, China
         \and
         Hebei Key Laboratory of High-Precision Computation and Application of Quantum Field Theory, Baoding, 071002,  China
         \and
         Hebei Research Center of the Basic Discipline for Computational Physics, Baoding, 071002, China
         }

   \date{Received ; accepted }

 
\abstract
   {
Analyzing filaments and cores in molecular clouds is key to understanding galactic star formation and its environmental dependence.
   This paper studies the properties and distribution of dense cores and filaments in the Ophiuchus molecular cloud, with a focus on the L1688 hub-filament system (HFS) and its star formation potential.
   We extracted sources and filaments from \textit{Herschel} images and a 13.5{\arcsec} resolution surface density map using the \textit{getsf} method, identified prestellar cores among the extracted sources, evaluated core mass segregation, and constructed the core mass function (CMF). We derived properties of the filaments from their radial surface density profiles, constructed the filament linear density function (FLDF), and assessed the mass distribution in the L1688 HFS to estimate the core and filament formation efficiencies (CFE, FFE).
   We identified 64 protostellar, 132 prestellar, and 686 unbound cores. The CMF of the prestellar cores has a power-law exponent of $-0.86$ and the FLDF of the densest filaments has a similar slope of $-0.97$, whereas the CMF of the unbound cores is found to be $-1.36$. Mass segregation is prominent among the most massive cores, with only slight differences between the bound and unbound cores. The low-mass unbound cores affect the overall spatial distribution. Among the 769 well-resolved filaments, we find a median half-maximum width of 0.12 pc and a median slope of $-1.4$ for the filament radial profiles. Mass distribution in the L1688 hub is dominated by the filaments and outside the hub it is dominated by the molecular cloud background. There exist a strong correlation between FFE and CFE that reach their respective maxima of 71\% and 5\% within the hub and decrease to 21\% and 0.9\% outside it.
    The results suggest that the gravitational potential in the L1688 HFS influences core clustering in its high-density regions and that the filament-dominated core formation is a key mechanism in star formation within the system.
    }

   \keywords{Stars: formation -- Infrared: ISM -- Submillimeter: ISM -- Methods: data analysis -- Techniques: image processing -- Techniques: photometric}

   \maketitle
%

\section{Introduction}           
\label{sect:intro}

Stars form within the densest areas of molecular clouds, known as dense molecular cores, where their self-gravity overcomes the gas pressure and all other supporting forces, thereby leading to their collapse and the formation of stars \citep{Williams+2000,BerginTafalla2007,Dib+2008,Andre+2014,HeyerDame+2015}. 

Over the last decade, filamentary structures of molecular clouds gained significant attention as the key sites of core formation, playing a crucial role in the process of star formation. Filaments are now considered to be ubiquitous in star-forming regions and are thought to channel material from the diffuse interstellar medium into the densest regions, where prestellar cores emerge and eventually form stars \citep{Andre+2010, Molinari+2010, Arzoumanian+2011, Zhang+2020,Ren+2023}. Analyses of observations with the \textit{Herschel} Space Observatory have shown that these filaments typically exhibit a characteristic width of $\sim$0.1 pc, across different environments and cloud types, suggesting a common formation mechanism \citep{Arzoumanian+2019,Andre+2022}. However, this typical width has been debated \citep{Panopoulou+2017, Panopoulou+2022}, as interferometric observations using dense gas tracers (e.g., $\mathrm{N_2H^+}$ and $\mathrm{NH_3}$) in regions such as Orion and Serpens South have revealed significantly narrower filament widths \citep{Fernandez+2014, Monsch+2018, Hacar+2018}.

Mass distribution of cores within filaments is described by the core mass function (CMF) which for some star-forming regions was shown to have a shape similar to that of the Galactic field stellar initial mass function (IMF), particularly at its high-mass end \citep{Motte+1998, Konyves+2015}. Understanding the connection between the CMF and IMF is essential, as the CMF likely represents the initial conditions for star formation. Despite extensive research, the processes governing star formation, especially the influence of molecular cloud structures on the distribution of core masses and the IMF, remain topics of active investigation \citep{Shu+1987,Krumholz2014,Dib+2023,Zhang+2024}. Numerous studies suggest that the fragmentation of filaments into prestellar cores is a hierarchical process driven by the interplay of gravitational instability and turbulence \citep{Padoan+Nordlund2002, Hennebelle+Chabrier2008}. However, understanding the detailed physical processes that link filaments, core formation, and their subsequent evolution into stars remain areas of ongoing research \citep{Andre2017}.

The Ophiuchus molecular cloud, at its distance $d{\,\approx\,}144$ pc \citep[][]{Zucker+2020}, is a perfect target for studying low-mass star formation, because of its proximity and rich population of prestellar and protostellar cores \citep{Wilking+2008, Andre+2014}.  A census of the starless, prestellar, and protostellar cores from the multiwavelength \textit{Herschel} Gould Belt Survey (HGBS) images was done by \citet{Ladjelate+2020} using the source and filament extraction methods \textit{getsources} \citep[][]{Menshchikov+2012} and \textit{getfilaments} 
\citep[][]{Menshchikov2013}. Previous studies, based on the \textit{Herschel} images, have revealed a complex filamentary network in Ophiuchus, with the L1688 region identified as a hub-filament system (HFS) where multiple filaments converge and drive star formation \citep{Ladjelate+2020}.
These dense filaments concentrate material and are conducive to core formation and star formation as shown by the tight core–filament correlation in Ophiuchus \citep{Ladjelate+2020}.

In this work, we used the new source and filament extraction method \textit{getsf}{\footnote{\url{https://irfu.cea.fr/Pisp/alexander.menshchikov/}}} \citep{Menshchikov2021method} with significantly improved algorithms over those used by \citet{Ladjelate+2020}. The \textit{getsf} method separates the structural components of sources, filaments, and backgrounds before extracting both cores and filaments within a consistent approach, which results in their more accurate detection and measurement \citep[][]{Menshchikov2021benchmark}. 
Benchmark tests \citep[][]{Menshchikov2021benchmark} show that \textit{getsf} achieves a completeness of about 60–70\% for core detection in complex backgrounds, with measurement uncertainties of 5–10\%. In simple backgrounds, the completeness exceeds 80\%. By comparison, \textit{getsources} reaches 50–70\% completeness and systematically underestimates source sizes by approximately 20\%. Furthermore, \textit{getsf} enables accurate extraction of filamentary structures, with width uncertainties of 10–20\%, whereas \textit{getsources} treated filaments as part of the background and failed to reconstruct their structures reliably. Crucially, the tools used by \citet{Ladjelate+2020} did not support radial filament measurements, making such analyses unfeasible. In contrast, \textit{getsf} allows us to quantitatively characterize filament radial profiles and derive new physical parameters. We also employed the maps of H$_2$ surface densities and dust temperatures created at a high 13.5{\arcsec} resolution with the \textit{hires} method \citep{Menshchikov2021method}, in contrast to the 18.2{\arcsec} resolution adopted by \citet{Ladjelate+2020}. The higher angular resolution enables the identification of finer structural details and a more accurate detection and measurement of both cores and filaments. The enhanced sensitivity and ability to detect faint, complex filamentary structures is paramount in refining the characteristics of the L1688 HFS, hence in understanding the mechanisms driving star formation and the role of these structures in regulating the star formation efficiency (SFE). Detailed studies of core mass segregation and filament alignment within the HFS are essential for advancing our knowledge of star formation processes in this region \citep{Dib+Henning2019, Kumar+2020}.

In Sect.~\ref{sec:herschel_science_archive_data}, we present the \textit{Herschel} dust continuum data for the Ophiuchus molecular cloud and construction of the high-resolution surface densities. Sect.~\ref{sec:data_analysis_results} describes the extraction of sources and filaments, the identification of reliable cores and filaments, their measured properties, core mass functions (CMF), and filament linear density functions (FLDF). Sect.~\ref{StructAnalysis} analyzes the L1688 hub morphology, spatial distribution and mass segregation of starless cores, radial structure of the hub, and core and filament formation efficiencies (CFE, FFE). In Sect.~\ref{discuss}, we discuss the implications of the CMF, FLDF, mass segregation, and their roles in the filament-driven core formation in the L1688 HFS. Key results and conclusions are summarized in Sect.~\ref{conc}.

\section{Observational Data}
\label{sec:herschel_science_archive_data}

The Ophiuchus molecular cloud is one of the regions mapped in the \textit{Herschel} Gould Belt Survey (HGBS) \citep{Andre+2010}. The observations were carried out using the Photodetector Array Camera and Spectrometer (PACS) \citep{Poglitsch+2010} and the Spectral and Photometric Imaging Receiver (SPIRE) \citep{Griffin+2010} in the parallel mode with a scanning speed of 60{\arcsec}s$^{-1}$. The resulting PACS data at 70 and 160\,$\mu$m have the angular resolutions of 8.4 and 13.5{\arcsec}, respectively, whereas the SPIRE data at 250, 350, and 500\,$\mu$m have the resolutions of 18.2, 24.9, and 36.3{\arcsec}, respectively.

The data products were retrieved from the \textit{Herschel} Science Archive (using the level 2.5 data)\footnote{\url{http://archives.esac.esa.int/hsa/whsa/}}. Specific observations for different parts of the Ophiuchus cloud include L1688, observed in September 2010 (IDs 1342205093 and 1342205094) and L1712 (IDs 1342204088 and 1342204089). For the Northern Streamer, the observations were done in February 2011 (IDs 1342214577 and 1342214578). An additional observation was conducted in March 2012 to address incomplete coverage in the PACS 70 and 160 $\mu$m bands at the overlapping areas of these three regions (IDs 1342241499 and 1342241500).

We used the \textit{montage} software\footnote{\url{http://montage.ipac.caltech.edu}} to stitch the level 2.5 images into a complete map of the Ophiuchus cloud (Fig.~\ref{Herscheloph}), applying background corrections at the overlapping edges to eliminate jagged noise structures. The Ophiuchus cloud covers an area of approximately 26\,deg$^2$ at 70 and 160\,$\mu$m, and approximately 27\,deg$^2$ at 250--500\,$\mu$m. Derivation of the high-resolution images of surface densities and temperatures (at 13.5{\arcsec}, Appendix~\ref{sec:highres_density_maps}) consisted of estimating zero offsets by comparing the \textit{Herschel} data with the \textit{Planck} data \citep[e.g.,][]{Bernard+2010, Bracco+2020} and pixel-by-pixel SED fitting using the \textit{hires} algorithm \citep{Menshchikov2021method}. Consistency of temperature measurements across different wavebands and Fourier analysis of surface density maps are discussed in Appendix~\ref{sec:highres_density_maps}.

\section{Data Analysis and Results}
\label{sec:data_analysis_results}

\subsection{Source and Filament Extraction}
\label{sourceextract}

\begin{figure*}
\centering
\includegraphics[width=\hsize]{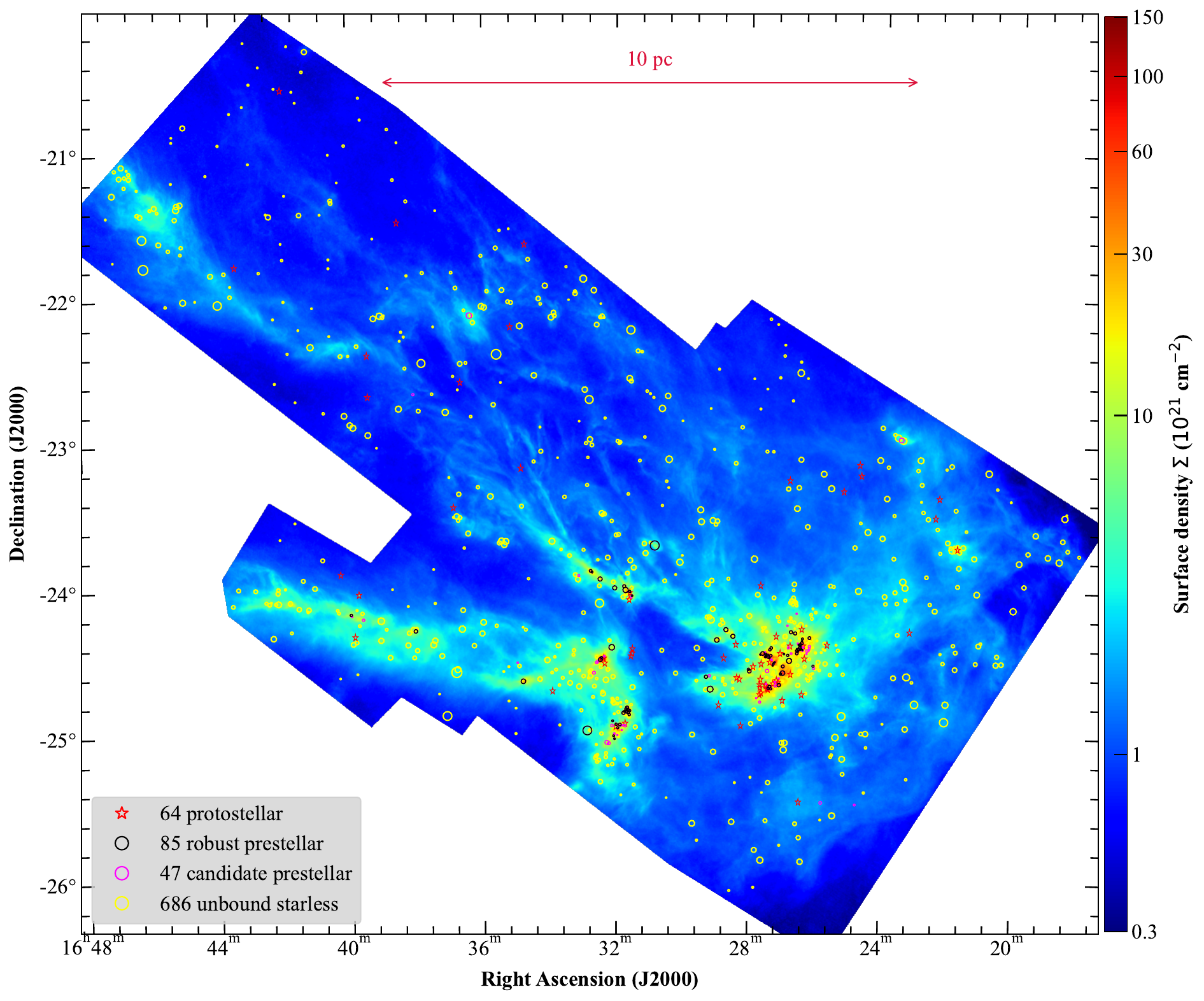}
\caption{Spatial distribution of the 882 dense cores in the Ophiuchus molecular cloud, extracted from the multiwavelength \textit{Herschel} images and the 13.5{\arcsec} resolution map of surface densities $\Sigma$. The starless cores are shown with their elliptical footprints measured in $\Sigma$, whereas the protostellar cores are marked with red pentagrams.}
\label{fig:ophiuchus_cores}
\end{figure*}

Sources and filaments were extracted with the multiscale, multiwavelength method \textit{getsf} \citep{Menshchikov2021method}. For the detection of both sources and filaments, we used the \textit{Herschel} images at 250, 350, and 500\,$\mu$m, along with the 13.5{\arcsec} surface density map. Sources were measured in all \textit{Herschel} images (70--500\,$\mu$m) and the surface densities, whereas filaments were measured only in the image of surface densities.

The single free parameter of \textit{getsf}, a maximum size of the structures to be extracted in an image, was visually constrained from the observed images as the radius of the footprint (full extent of the structure) of the largest source or filament of interest. A value 4 times the maximum size sets practical upper limits on the largest scales to be processed in the spatially decomposed images during separation of structural components and detection of sources and filaments \citep[Sect.~3.1.3 in][]{Menshchikov2021method}. We adopted the source maximum sizes of 13, 90, 120, 160, and 240{\arcsec} for the images at 70--500\,$\mu$m, respectively, and 90{\arcsec} for the 13.5{\arcsec} resolution map of surface densities. The sizes (widths) of filaments of interest were set to 15, 110, 150, 200, and 300{\arcsec} for the images and 110{\arcsec} for the surface densities.

Observed peak emission of the relatively hot protostars and their parent cores (with colder dust) can appear at slightly different positions across different wavelengths, mainly because of the differences in angular resolutions. Protostars are usually identified by their emission at shorter wavelengths, such as 70\,$\mu$m, where the internal heating by accretion energy makes them more prominent \citep{Andre+2010, Hennemann+2010}. Accordingly, we separated protostars from starless cores by detecting protostars in the 70\,$\mu$m \textit{Herschel} image \citep{Konyves+2015}.

\subsection{Selection of Candidate Cores}
\label{selectcores}

Sources are defined in \textit{getsf} as the emission peaks with relatively circular intensity profiles, that are significantly stronger than the local background and noise fluctuations \citep{Menshchikov+2012}. In total, \textit{getsf} extracted 2758 sources. To remove possible spurious detections and have only the reliable and well-measurable sources, we applied the basic (weak) selection criteria described in \cite{Menshchikov2021benchmark}. For each acceptably good source, we required that the signal-to-noise ratios $F_{\text{T}\,} \sigma^{\,-1}_{\text{T}} > 1$ (for the integrated flux) and $F_{\text{P}\,} \sigma^{\,-1}_{\text{P}} > 1$ (for the peak intensity), and that the monochromatic detection significance $\Xi > 1$ and goodness $\Gamma > 1$. These criteria were applied to the sources measured at 160--500\,$\mu$m, and in the 13.5{\arcsec} surface densities. 

\begin{figure}
       \centering
       \includegraphics[width=1.0 \hsize]{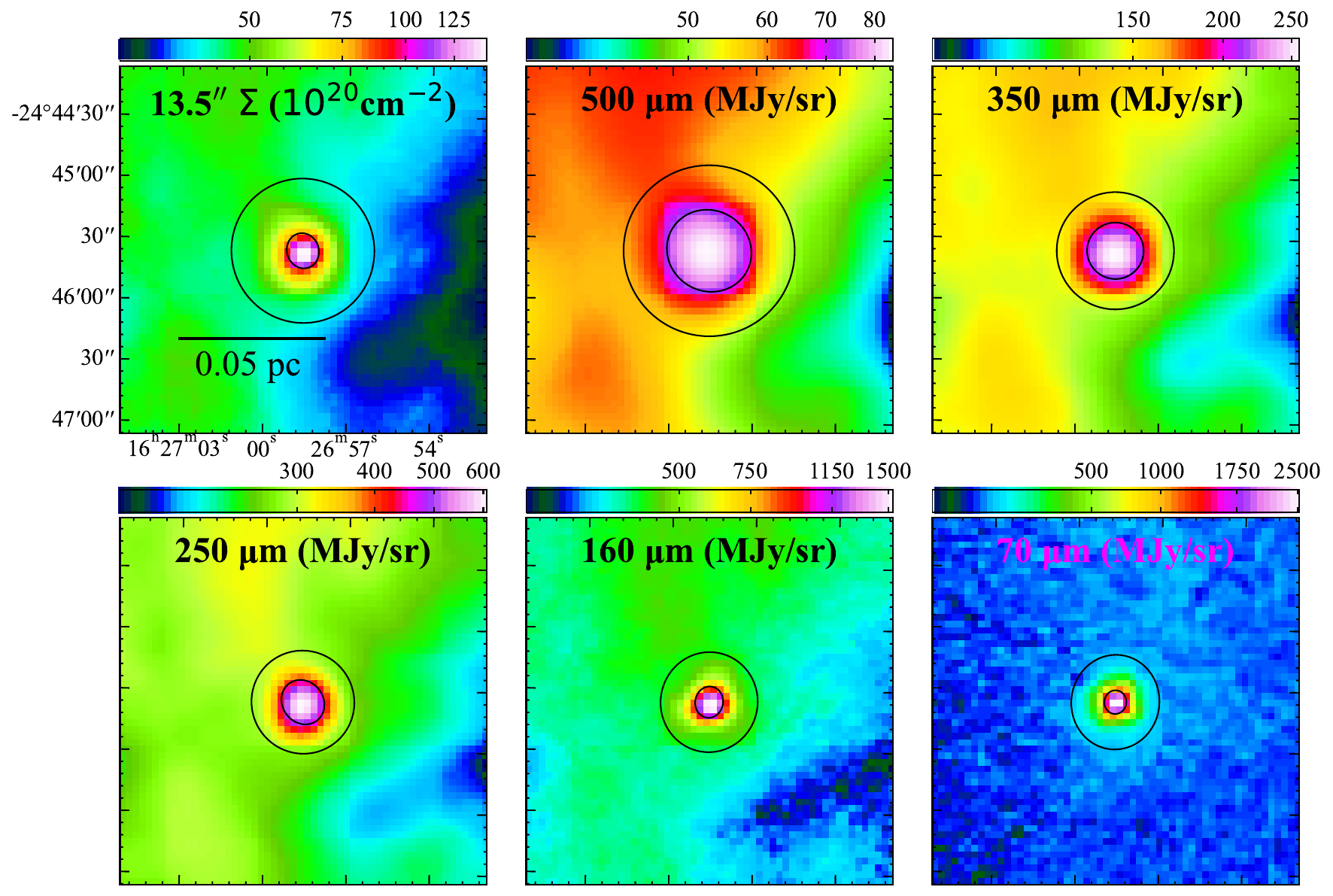}
       \caption{Example of the images of a protostellar core in five \textit{Herschel} wavebands and surface density $\Sigma$. The inner and outer ellipses visualize the FWHM sizes and the footprint sizes, respectively (cf. Table \ref{coreobs}).}
       \label{fig:protostellar}
\end{figure}

We also employed stronger selection criteria \citep[see Eq.~(1) in][]{Menshchikov2021benchmark}, based on benchmarking results. They require that $F_{\text{T}\,} \sigma^{\,-1}_{\text{T}} > 2$ and $F_{\text{P}\,} \sigma^{\,-1}_{\text{P}} > 2$, the ratio of the major and minor half-maximum sizes $A {B^{-1}} < 2$ (source not too elongated), and the major footprint size obeys $A_\text{F\,} {A^{-1}} > 1.15$ (footprint not too small). A source was considered a candidate core if it satisfied the basic selection criteria in at least two different wavebands and in the 13.5{\arcsec} surface densities, as well as the benchmark selection criteria in at least one waveband. With the above criteria, we identified 882 candidate cores, displayed in Fig.~\ref{fig:ophiuchus_cores}. Individual images of each candidate core in each band and in the 13.5{\arcsec} surface density map have also been used for further detailed examination, as illustrated in Fig.~\ref{fig:protostellar}.

Following the methodology employed in the HGBS studies \citep[e.g.,][]{Konyves+2015}, we performed an additional source extraction in the 70\,$\mu$m image and identified protostars by requiring that the sources exhibit narrow peaks with a relatively round shape: their average half-maximum sizes $H = (AB)^{1/2} < 1.5 \times 8.4{\arcsec}$ and major to minor size ratio $AB^{-1} < 1.3$. To further refine the sample, we cross-referenced these sources with the SIMBAD database\footnote{\url{https://simbad.cds.unistra.fr/simbad/}} and the NASA/IPAC extragalactic database (NED)\footnote{\url{https://ned.ipac.caltech.edu}}, excluding galaxies and other non-stellar objects. 

\subsection{Nature of Extracted Cores}
\label{classification}

Starless cores are cold, dense regions in space, devoid of significant infrared emission at wavelengths $\lambda \la 160$ $\mu$m because of the absence of internal heat sources; they represent the earliest stage in star formation. Such cores are further categorized as the gravitationally bound, prestellar cores, that are likely to collapse and form stars, and unbound starless cores, which can disperse with time \citep{Alves+2001, BerginTafalla2007, WardThompson+2007, Andre+2014}.

Protostellar cores represent a more advanced stage than the starless cores in the star formation process, because they contain an accreting stellar embryo or protostar at their centers \citep{Francesco+2007, Evans+2009}. Gas and dust in these cores are actively accreted by the protostar that gradually accumulates its mass \citep{Andre+2000, Dunham+2014}. In our analysis, the candidate cores are classified as the protostellar ones, if they contain a protostar within them, manifested by a detectable and measurable peak at 70 $\mu$m (Fig.~\ref{fig:protostellar}). We found 64 protostellar cores.

\begin{figure}
  \centering
  \includegraphics[width=1.0\hsize]{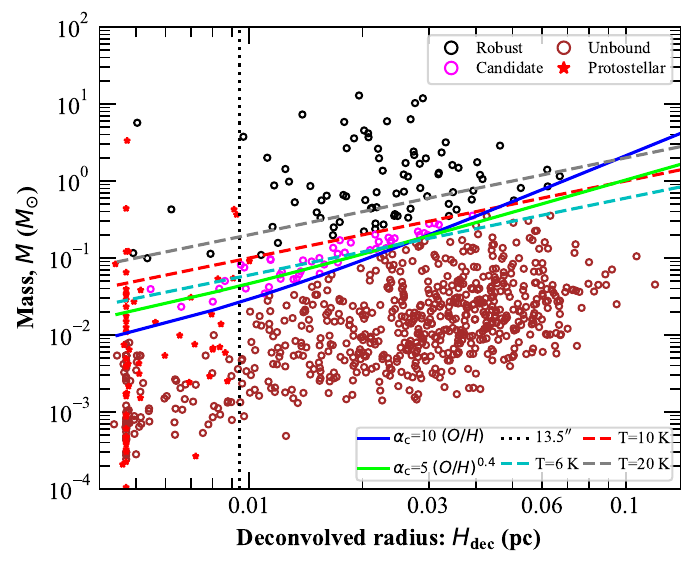}
  \caption{Mass-size diagram for dense cores in the Ophiuchus molecular cloud. The diagram classifies starless cores into robust prestellar, candidate prestellar, and unbound starless categories based on their positions relative to critical stability lines. The vertical dotted line marks a scale of 0.009 pc, corresponding to the 13.5{\arcsec} beam size. Overlaid dashed curves represent the critical Bonnor-Ebert spheres at temperatures of 6, 10, and 20 K, with the 10 K curve used to identify robust prestellar cores. The blue solid curve shows the empirical relation $\alpha_{\rm c} = 10\,(O/H)$ for selecting candidate prestellar cores (see Sect.~\ref{classification}), while the green solid curve represents the empirical relation $5\,(O/H)^{0.4}$ used by \citet{Konyves+2015}. Also shown are the protostellar cores with steeper temperature and density distributions are mostly unresolved \citep{Li+2023}.}
  \label{fig:core_mass_size_diagram_ophiuchus}  
\end{figure}

The stability of the prestellar cores is often determined using the Bonnor–Ebert (BE) model, which describes the equilibrium between gas pressure and self-gravity in isothermal, hydrostatic spheres \citep{Ebert1955, Bonnor1956}. Although this is a simplified model that neglects turbulence \citep{Galli+2002, Ballesterosparedes+2003, Dib+2007, Li+2013} and magnetic fields \citep{Dib+2010}, it is widely used to approximately estimate the mass sufficient for the onset of core collapse. The critical mass can be expressed as $M_{\rm BE} \approx 2.4 R_{\rm BE\,} c_{\rm s}^2 / G$, where $c_{\rm s}$ is the sound speed, $R_{\rm BE}$ the BE sphere radius, and $G$ the gravitational constant. We adopted the sound speed $c_{\rm s}$ for a temperature of 10 K and approximated $R_{\rm BE}$ by the deconvolved half-maximum size $\tilde{H} = (H^2 - O^2)^{1/2}$, where $H = (AB)^{1/2}$ is the geometric mean of the major and minor half-maximum sizes $A$ and $B$ of the cores from their surface densities (resolution $O = 13.5{\arcsec})$ and the approximation $R_{\rm BE} \approx \tilde{H}$ is valid to within 20\% \citep[Fig.~4 in][]{Menshchikov2023}. To avoid large errors, we deconvolved only the partially-resolved and resolved cores with $H > 1.1\,O$, whereas for the unresolved cores with $H \leq 1.1\,O$ we arbitrarily adopted $\tilde{H} = 0.5\,H$ \citep{Menshchikov2023}.

Masses $M$ of the cores were derived by fitting their spectral energy distributions (SED), using the background-subtracted fluxes $F_{\rm T}$ measured in the 160, 250, 350, and 500 $\mu$m wavebands. We emloyed the \textit{fitfluxes} utility from the \textit{getsf} software, applying a modified optically-thin blackbody model \citep[\textit{thinbody},][]{Men2016fitfluxes}. We adopted the distance $d = 144$ pc to the Ophiuchus molecular cloud and the dust opacity $\kappa = \kappa_0 \left(\nu/\nu_{0}\right)^{\,\beta}$, where $\nu$ is the frequency, $\kappa_0 = 0.1$ cm$^2$ g$^{-1}$ (per gram of dusty gas), $\nu_{0} = 10^3$ GHz, and $\beta = 2$. The fitting provided the estimated core masses $M$, mass-averaged dust temperatures $T$, and the associated uncertainties. It is necessary to emphasize that the mass uncertainties are usually large \citep[at least a factor of 2--3,][]{Men2016fitfluxes}.

We denote starless cores with the ratio $\alpha_{\rm BE} = M_{\rm BE} / M \leq 2$ as the ``robust'' prestellar cores, because the Bonnor-Ebert sphere represents an equilibrium solution, where the gravitational stability is more stringent than the virial mass condition for a core to be gravitationally bound \citep{Li+2013}. We identified 85 robust prestellar cores. Furthermore, studies of the Aquila and California molecular clouds \citep{Konyves+2015, Zhang+2024} suggested that using the $\alpha_{\rm BE} \leq 2$ condition might be overly conservative for identifying the gravitationally bound cores. Following \citet{Zhang+2024}, we adopted the modified empirical critical value $\alpha_{\rm c} = 10\,(O/H)$ that is inversely proportional to the core resolvedness \citep[defined as $H/O$,][]{Menshchikov2023} in surface densities. With the additional condition $2 < \alpha_{\rm BE} \leq \alpha_{\rm c}$, we identified 47 candidate prestellar cores. The remaining 686 unbound starless cores (with $\alpha_{\rm BE} > \alpha_{\rm c}$) are unlikely to collapse and form stars. The spatial distribution of all identified cores and the core mass-size diagram are presented in Figs.~\ref{fig:ophiuchus_cores} and \ref{fig:core_mass_size_diagram_ophiuchus}, respectively.

\begin{figure}
    \centering
    \includegraphics[width=1.0\hsize]{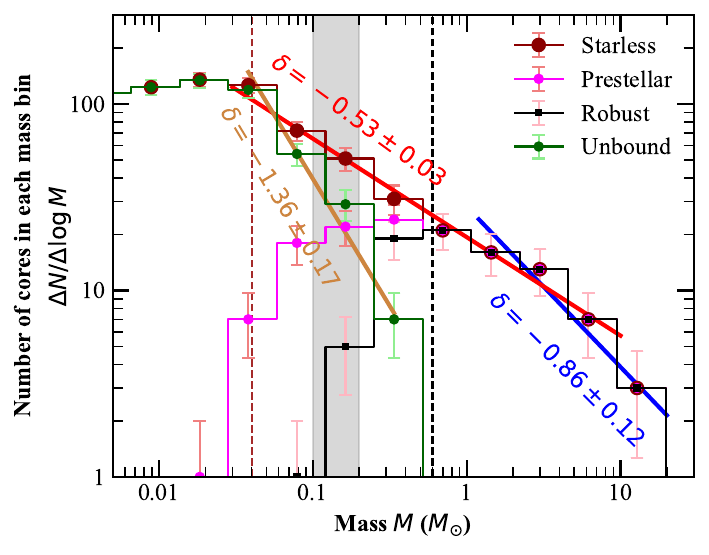}
    \caption{CMFs for the starless cores and the subsets of prestellar (candidate and robust) cores in the Ophiuchus molecular cloud. The CMF of the entire population of starless cores can be described by a power law with $\delta = -0.53 \pm 0.03$, whereas for the unbound cores it follows a power law with $\delta = -1.36 \pm 0.17$ and for the robust prestellar cores it has a significantly shallower slope $\delta = -0.86 \pm 0.12$ (at $M > 1 M_{\sun}$). The gray shaded area includes the limiting masses $M_{0}$ from the completeness modeling in three star-forming regions observed with \textit{Herschel}: California \citep{Zhang+2024}, Aquila \citep{Konyves+2015}, and Orion B \citep{Konyves+2020}, scaled to the distance of the Ophiuchus cloud. The extracted cores may be relatively complete (Sect.~\ref{coremassfun}) at $M_{0} \gtrsim 0.6 M_{\sun}$ for the prestellar cores (black dashed line) and at $M_{0} \gtrsim 0.04 M_{\sun}$ for the entire ensemble of starless cores (brown dashed line).}
    \label{fig:CMF}  
\end{figure}

\subsection{Mass Functions of the Cores}
\label{coremassfun}

The CMF is defined as the number of cores per unit mass interval and its high-mass part is usually approximated by ${\rm d}N/{\rm d}\log M \propto M^{\,\delta}$. This form is compatible with the Salpeter IMF for stars ($\delta = -1.35$) and it captures the observed power-law distribution of core masses in molecular clouds \citep{Salpeter1955, Motte+1998, Johnstone+2000, Dib+2008, Ballesteros-Paredes+2020}. The CMF of the entire sample of starless cores in the Ophiuchus molecular cloud displays a slope $\delta = -0.53 \pm 0.03$ for the masses in the range 0.04--10\,$M_{\sun}$ (Fig.~\ref{fig:CMF}), much shallower than that of the Salpeter IMF. However, the most massive prestellar cores of 1--20\,$M_{\sun}$ reveal a substantially steeper CMF with $\delta = -0.86 \pm 0.12$, whereas the sample of unbound cores shows $\delta = -1.36 \pm 0.17$, essentially the Salpeter slope.

To better understand these results, we spatially separated the CMFs for two regions of the Ophiuchus molecular cloud, above and below a certain surface density level of the cloud. With a core background value $\Sigma_{\rm D} \approx 2\times 10^{22}$\,cm$^{-2}$, chosen as the level, dividing the high- and low-density regions, most of the robust prestellar cores are found in the hub and densest filaments outside it, whereas most of the unbound starless cores are located in the lower-density area of the map in Fig.~\ref{fig:ophiuchus_cores}. In a range of surface densities within a factor of two above and below $\Sigma_{\rm D}$, relatively small fractions of the unbound and bound cores co-exist with the candidate prestellar cores. The spatially separated CMFs are very similar to those in Fig.~\ref{fig:CMF}, therefore they are not presented here. 

The steep CMF of the unbound starless cores suggests that the cores represent low-background density enhancements of the Ophiuchus molecular cloud. Indeed, \textit{Herschel} observations clearly demonstrated that such interstellar clouds spatially fluctuate quite significantly on all scales. On the other hand, the shallow CMF slopes of the prestellar cores must be related to their formation in the high-density areas. The shallow CMF of prestellar cores in the Ophiuchus cloud is similar to that presented by \cite{Ladjelate+2020}. It is relevant to note that recent ALMA-IMF observations \citep{Pouteau+2022,Louvet+2024} also found a similar slope ($\delta = -0.97$) for dense regions of high-mass star formation.

An important uncertainty, implicitly present in the astrophysical interpretations of CMFs, is that the masses derived by the SED fitting of integrated fluxes, may be insufficiently accurate to make reliable conclusions. The biases and wide ranges of errors associated with the masses \citep{Men2016fitfluxes} can significantly redistribute the cores between the mass bins and distort the shape of an observationally determined mass function with respect to the true CMF of the observed objects.

When analyzing and interpreting derived CMFs, observational studies often employ simulated images populated with radiative-transfer models of sources and/or filaments. Extractions in such images allow to judge how complete the extracted set of dense cores can be and below what limiting mass $M_{0}$ the fraction of extracted cores or filaments starts to rapidly drop. However, it is a non-trivial problem to construct the simulated images closely resembling the observed set of images in each waveband, especially for the \textit{Herschel} images with their bright (dense), highly-structured filamentary backgrounds \citep{Men2016fitfluxes,Menshchikov2023}.

In the absence of a satisfactory, accurate solution of the problem, we decided not to perform the simulations for completeness evaluation in this work. However, we scaled the limiting masses $M_{0}$ of prestellar cores, obtained in previous studies of nearby star-forming regions, to the distance of the Ophiuchus molecular cloud to see, whether its CMF (Fig.~\ref{fig:CMF}) is compatible with the previously published CMFs. The scaled values turned out to be fairly consistent with each other: for the California region \citep{Zhang+2024} it scales to $M_{0} = 0.1{-}0.2$ $M_{\sun}$, for the Aquila region \citep{Konyves+2015} to $M_{0} = 0.1$ $M_{\sun}$, and for the Orion B region \citep{Konyves+2020} to $M_{0} = 0.14$ $M_{\sun}$. Taking into account that the values may be underestimated by roughly a factor of $\sim$\,2--3, we presume that they are likely to point to $M_{0}\approx$ 0.4--0.6 $M_{\sun}$ for prestellar cores for the Ophiuchus cloud. The value is in the mass bin (Fig.~\ref{fig:CMF}), where the CMF starts to deviate from the high-mass power law and to morph into a shape reminiscent of the log-normal curve. Similarly, the CMF of the unbound starless cores deviates from the power law at much lower masses $M < 0.04$\,$M_{\sun}$, where our population of extracted starless cores becomes incomplete.

\begin{figure*}
     \centering
     \includegraphics[width=0.55\hsize]{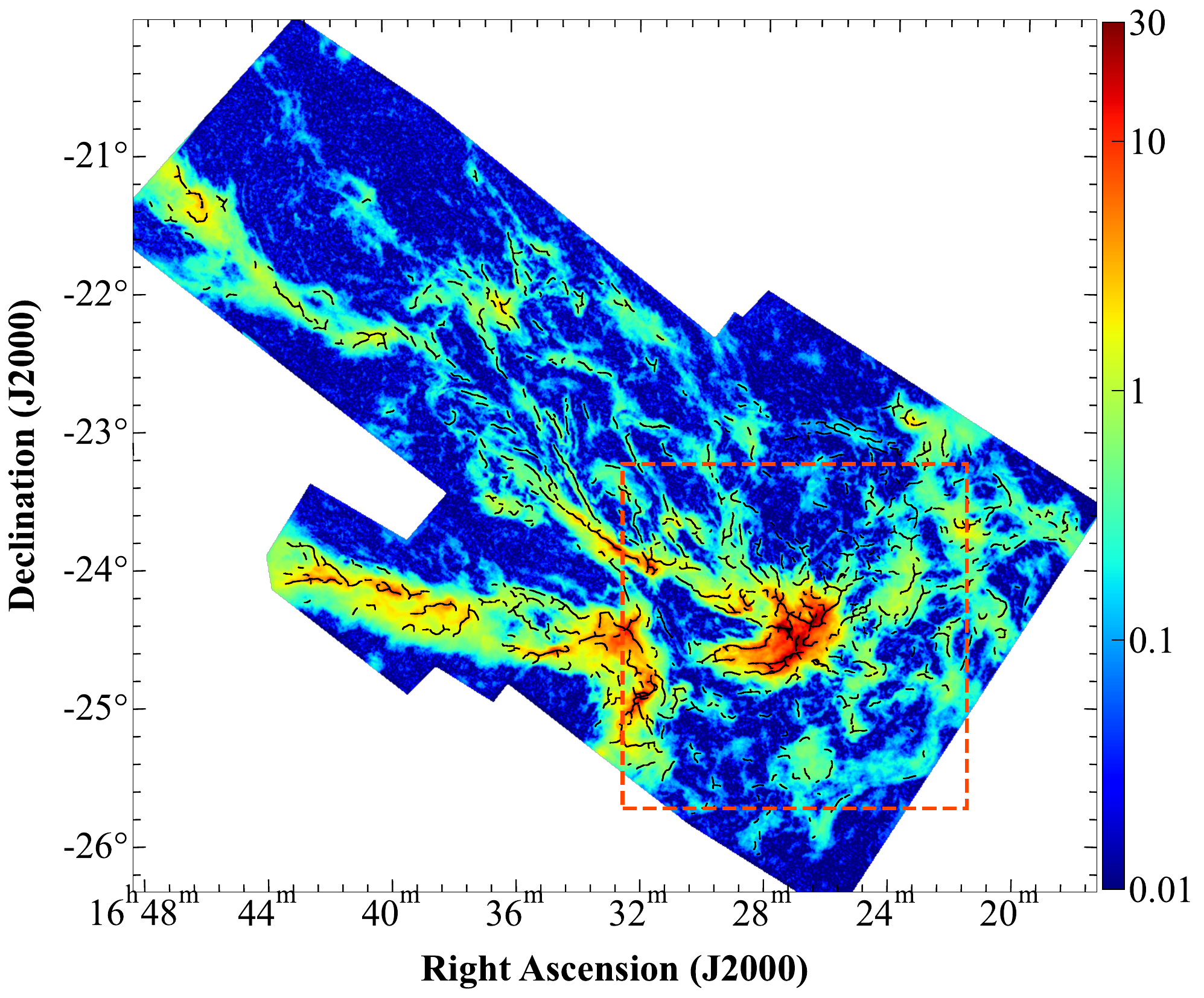}
     \includegraphics[width=0.43\hsize]{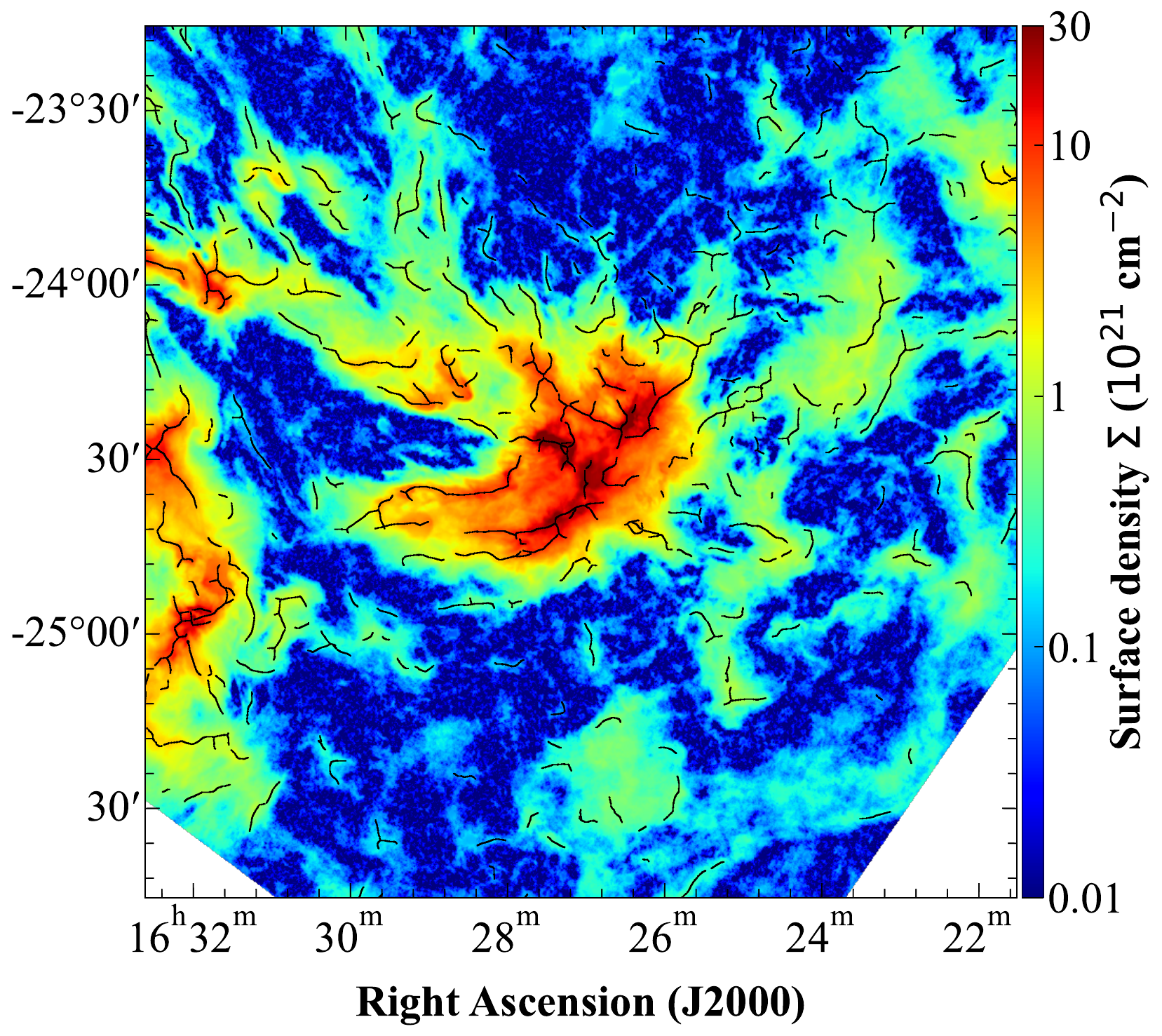}
     \caption{Filaments in the Ophiuchus molecular cloud (left panel) and the L1688 hub area (right panel). Shown is the image of the component of filamentary structures that are most prominent on spatial scales around 110{\arcsec}, separated by \textit{getsf} from both the sources and background cloud. The filament crests detectable across at least 5 consecutive scales (a factor of 1.3) are shown by the skeletons (black curves). Dashed rectangle in the left panel marks the area shown in the right panel.}
     \label{fig:1688C}  
\end{figure*}

\subsection{Filaments and Their Properties}
\label{filamentprops}

Filaments in molecular clouds are significantly elongated structures (e.g., Fig.~\ref{fig:1688C}) that are thought to play a fundamental role in the process of star formation \citep{Menshchikov+2010,Andre+2014,Zhang+2024}. Observations of the nearby molecular clouds with \textit{Herschel} suggested that widths $W$ of the filaments are distributed in a relatively narrow range around $\sim$ 0.1 pc \citep{Arzoumanian+2011,Arzoumanian+2019}. Filaments are believed to form via an interplay of various physical processes that include supersonic turbulence, gravitational collapse, and magnetic fields \citep[cf.][]{Andre2017}. Concentrations of gas and dust within the filaments lead to the formation of prestellar cores that eventually collapse and form stars \citep{Konyves+2015, Zhang+2020}. 

Accurate detection of the filaments is complicated by the fact that they are blended with other filaments and various nearby structures on the complex backgrounds in the \textit{Herschel} images. Measurements of the filament properties are also made inaccurate by background subtraction, when their true background is unknown (complex) and can only be guessed. Further difficulties are caused by the significantly nonuniform angular resolutions in the far-infrared wavebands. Images become much less sharp at longer wavelengths with lower angular resolutions, which aggravates the problems in distinguishing overlapping and intertwined filaments \citep{Menshchikov2023}. Moreover, filaments are three-dimensional structures that are interpreted on the basis of their observed two-dimensional projections.

Filamentary structures are observed on quite different spatial (angular) scales \citep[cf. Fig.~13 in][]{Menshchikov2021method}. Filaments in the Ophiuchus molecular cloud were extracted using \textit{getsf}, simultaneously with the source extraction. In this paper, we analyzed the filaments that are most prominent and detectable around spatial scales 110{\arcsec} (corresponding to the filament widths of 0.08\,pc) (Fig.~\ref{fig:1688C}). To exclude spurious detections, we selected only those filaments whose skeletons are traceable in at least 5 consecutive spatial scales (a factor of 1.3 in the scales). \noindent Some visually obvious but faint filaments were not detected (e.g., Fig.~\ref{fig:1688C}), primarily because the \textit{getsf} algorithm employs a multiscale analysis combined with stringent signal-to-noise criteria when extracting filament skeletons. Although such filaments appear visible in the filament component map, they do not show up in the final skeleton map \citep{Menshchikov2021method}. At each individual spatial scale, \textit{getsf} applies a cleaning threshold: only signal peaks exceeding the local background noise by approximately $2\sigma$ are retained as candidate filaments. If a faint filament falls below this threshold at any given scale, it is treated as noise and removed, thereby weakening its continuity and significance across multiple scales. To simplify the complex network of detected skeletons, \textit{getsf} eliminates their intersections, thereby creating a non-branching set of skeletons tracing the ``elementary filaments''. Surface density measurements for each filament were done in the images, where the sources had been removed and large-scale backgrounds subtracted. Radial density profiles were taken along the normals to the filament skeletons and, to ensure reliable measurements, only sufficiently isolated and well-resolved filaments were selected. 

\begin{figure}
  \centering
  \includegraphics[width=1.0\hsize]{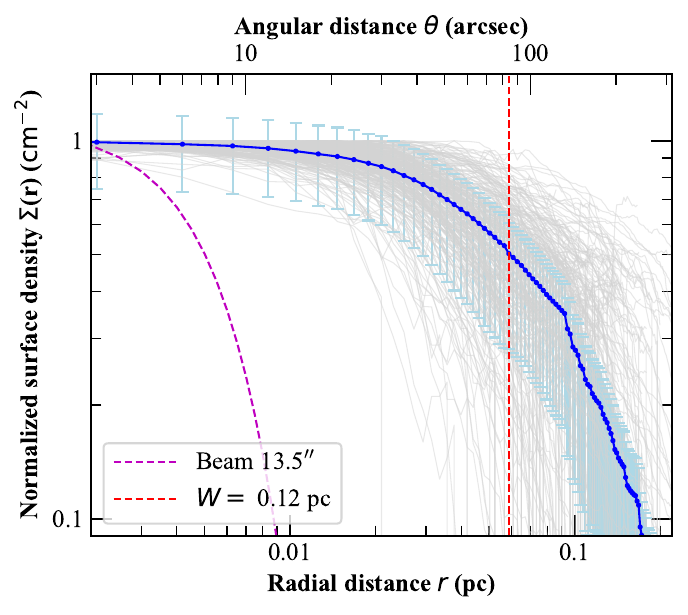}
   \caption{Normalized surface density profiles $\Sigma (r)$ of well-resolved filaments in the Ophiuchus molecular cloud, averaged over 769 measurable filaments (gray curves). Vertical bars show the range of profile variations along the filament crests. Dashed vertical line indicates the half-maximum radius of the average profile.}
  \label{fig:OphiuchusFilamentProfile}
\end{figure}

Following \cite{Zhang+2024}, we deemed a one-sided profile of a filament acceptably good, if the profile on that side extended to values below its half-maximum and the width was narrower than twice the width determined from the opposite side. To exclude the profiles contaminated by the background fluctuations or blending with other nearby filaments, we considered only the one-sided profiles that met these requirements. If both sides of a filament profile were acceptably good, the width $W$ was estimated as the arithmetic average of the one-sided median half-maximum widths $W_{\rm A}$ and $W_{\rm B}$. With this approach, we identified 769 filaments with measurable widths, whose average profiles are displayed in Fig.~\ref{fig:OphiuchusFilamentProfile}.

\begin{figure*}
   \centering
   \includegraphics[width=0.44\hsize]{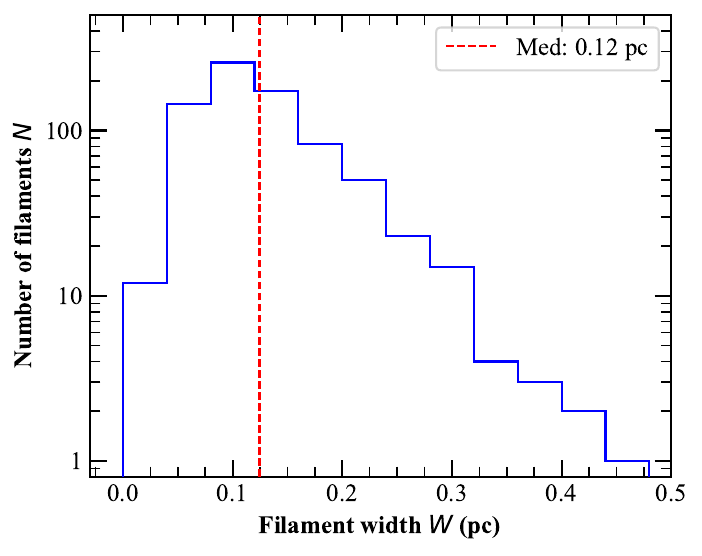}
   \includegraphics[width=0.44\hsize]{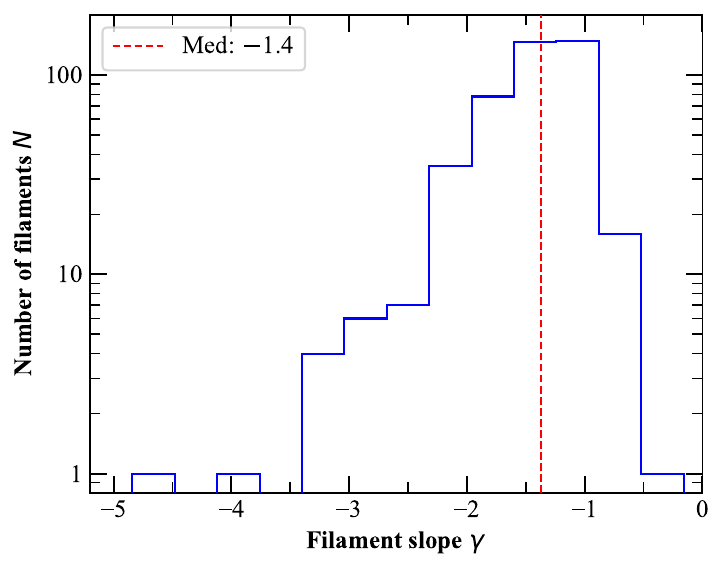}
   \caption{Distributions of filament widths $W$ (left panel) and profile slopes $\gamma$ (right panel) in the Ophiuchus molecular cloud for 769 filaments with measurable widths and 443 filaments with measurable slopes. Dashed lines indicate the median values of $W$ and $\gamma$.}
   \label{fig:WidthDistribution}
\end{figure*}

\begin{figure*}
  \centering
  \includegraphics[width=0.455\hsize]{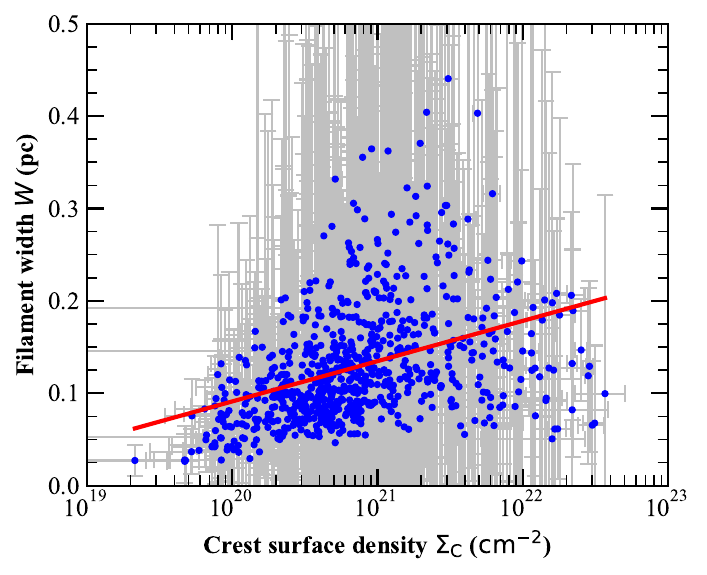}
  \includegraphics[width=0.455\hsize]{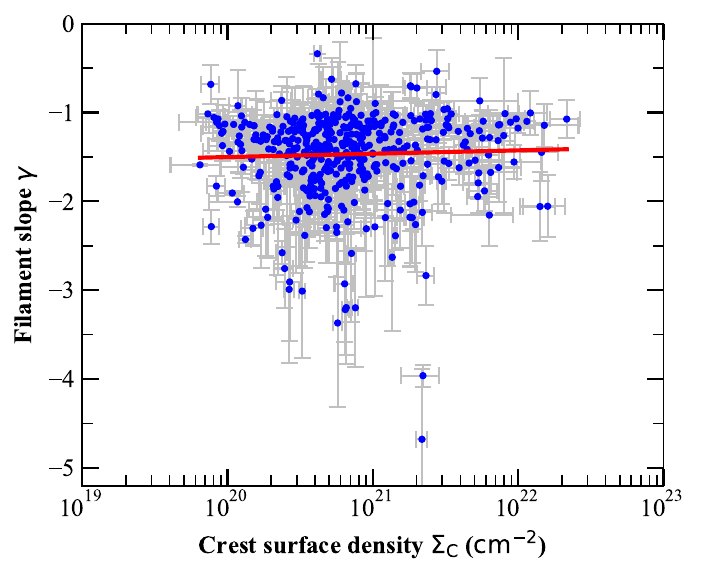}
   \caption{Dependence of the filament widths $W$ (left panel) and profile slopes $\gamma$ (right panel) on the crest surface density $\Sigma_{\mathrm{C}}$ for the 769 filaments with measurable widths and 443 filaments with measurable slopes. The red lines show $W = 0.0436\log\Sigma_{\rm{C}} - 0.78$ and $\gamma = 0.0377\log\Sigma_{\rm{C}} - 2.25$.}
  \label{fig:surfaceVSwidth}
\end{figure*}

In our analysis, the filament widths $W$ and crest surface densities $\Sigma_{\rm C}$ refer to the values averaged over the entire filament length. The widths are distributed in a range $0.02 \la W \la 0.4$ pc and become exponentially less abundant beyond a median width of 0.12 pc (Fig.~\ref{fig:WidthDistribution}). On average, the filaments have a tendency to have larger $W$ (by a factor of 4), when $\Sigma_{\rm C}$ increases by three orders of magnitude (Fig.~\ref{fig:surfaceVSwidth}). Although the median width is consistent with that found by \cite{Arzoumanian+2011,Arzoumanian+2019}, the agreement should not be interpreted as the confirmation of the previous findings of the quasi-universal width of filaments in star-forming regions. It is rather the consequence of our choice of the size of filaments of interest (110{\arcsec}) in the \textit{getsf} filament extraction. An investigation of the dependence of the filament properties on spatial scales will be done in our next paper. 

\begin{figure*}
   \centering
   \includegraphics[width=1.0\hsize]{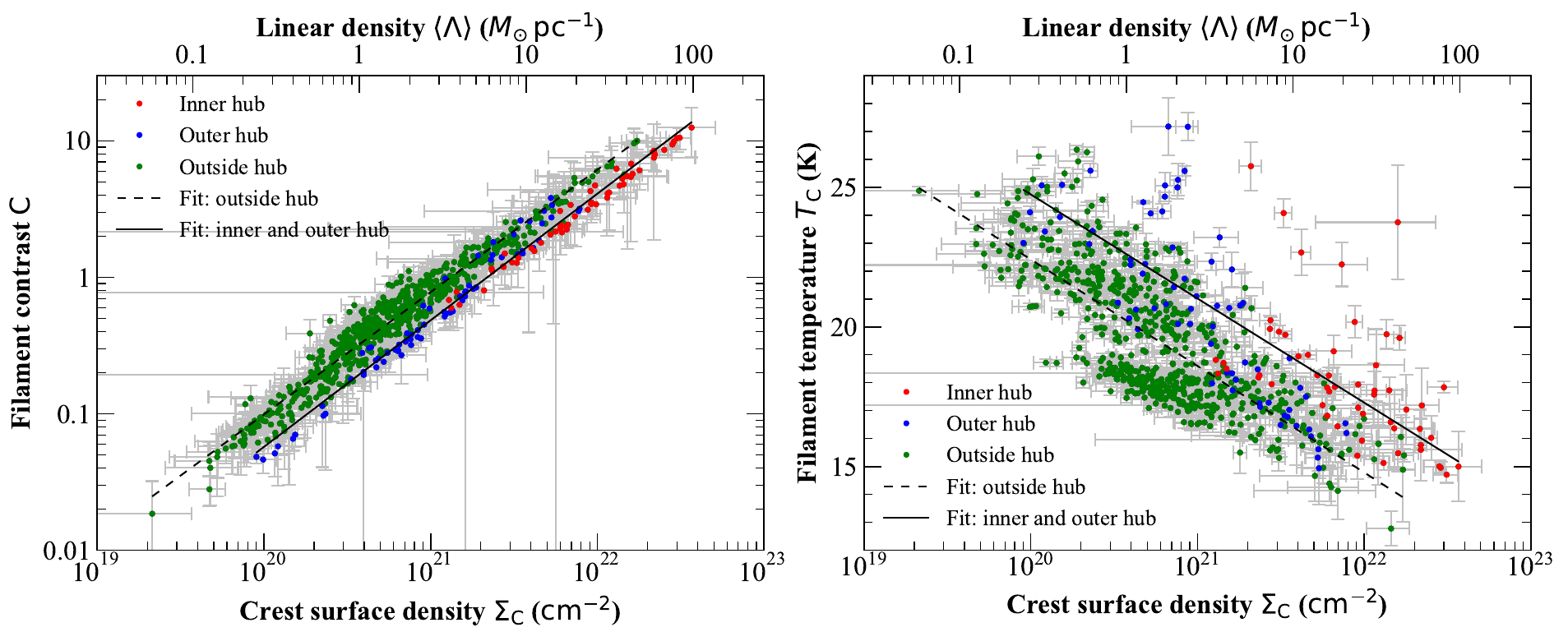}
\caption{Dependence of filament contrasts $C$ (left panel) and filament crest temperatures $T_{\rm C}$ (right panel) on the crest surface density $\Sigma_{\rm C}$ or representative linear density $\langle \Lambda \rangle$ for the filaments detected on spatial scales around 110$\arcsec$. The  properties of filaments in the inner hub area ($0 < r \le 0.85$ pc), outer hub area ($0.85 < r \le 1.45$ pc), and outside the hub ($r > 1.45$ pc) are plotted individually. The data can be represented by $C = 10^{-19.8\,}\Sigma^{0.93}_{\rm C}$, $T_{\rm C} = \log \Sigma^{-3.7}_{\rm C} + 99.3$ in the hub area (solid black lines) and $C = 10^{-18.9\,}\Sigma^{0.9}_{\rm C}$, $T_{\rm C} = \log\Sigma^{-3.8}_{\rm C} + 98.8$ outside the hub (dashed black lines). Error bars show the standard deviations of $C$, $T_{\rm C}$, and $\Sigma_{\rm C}$ along the filaments.}
   \label{fig:Nh2vsC}
\end{figure*}

Slopes of the filament radial profiles $\Sigma(r) \propto r^{\,\gamma}$ are defined as $\gamma = {\rm d}\ln \Sigma / {\rm d}\ln r$. In practice, we evaluated the one-sided slopes in the range $0.3 \leq \Sigma / \Sigma_{\rm C} \leq 0.6$ to exclude the inner flattened parts of the profiles, as well as their much fainter segments that are increasingly affected by the spatial fluctuations of surface density of the molecular cloud and inaccuracies of background subtraction. Only the slopes for the filaments with acceptably good one-sided widths were evaluated. If both sides of a filament had acceptable widths, then we adopted an arithmetically averaged slope. With this approach, we identified 443 filaments with measurable slopes, distributed in a relatively wide range $-3.4 \la \gamma \la -0.6$ with a median value of $-1.4$ (Fig.~\ref{fig:WidthDistribution}). The surface density slopes correspond to the volume densities profiles $\rho(r) \propto r^{\,\eta}$ with $-4.4 \la \eta \la -1.6$ and a median value of $-2.4$, similar to that found by \cite{Arzoumanian+2011,Arzoumanian+2019}. The filament slopes are practically invariant with the crest surface densities (Fig.~\ref{fig:surfaceVSwidth}).

It is useful to define contrasts of filaments as $C = \Sigma_{\rm C\,} / ^{\,}\Sigma_{\rm B}$, where $\Sigma_{\rm B}$ is the average background surface density of the filaments (along their skeletons).
We also can define a representative average linear density of the set of filaments, used for illustration purposes in our paper, as $\langle \Lambda \rangle = \mu_{\rm H_{2}\,} m_{\rm H\,}\Sigma_{\rm C\,} \langle W\rangle$, where $\mu_{\rm H_{2}}\, (= 2.8)$ is the mean molecular weight of gas per H$_2$ molecule and $m_{\rm H}$ is the hydrogen mass. Relationship between the the filament contrast and the surface density $\Sigma_{\rm C}$ or representative linear density $\langle \Lambda \rangle$ is displayed in Fig.~\ref{fig:Nh2vsC}. The data show a clear positive correlation between $C$ and both $\Sigma_{\rm C}$ density and linear density, which is consistent with the findings by \cite{Zhang+2024}. As filaments accumulate more material and their linear density increases, they become more distinct from their surroundings, which implies higher contrast values. Figure~\ref{fig:Nh2vsC} shows also an inverse relationship between the average crest dust temperatures $T_{\rm C}$ and the surface (or linear) densities. The crest temperature decreases by roughly 10 K as $\Sigma_{\rm C}$ increases by three orders of magnitude. As expected, the denser filaments shield their interiors from external radiation more efficiently, which leads to lower dust temperatures.

\begin{figure}
    \centering
    \includegraphics[width=1.0\hsize]{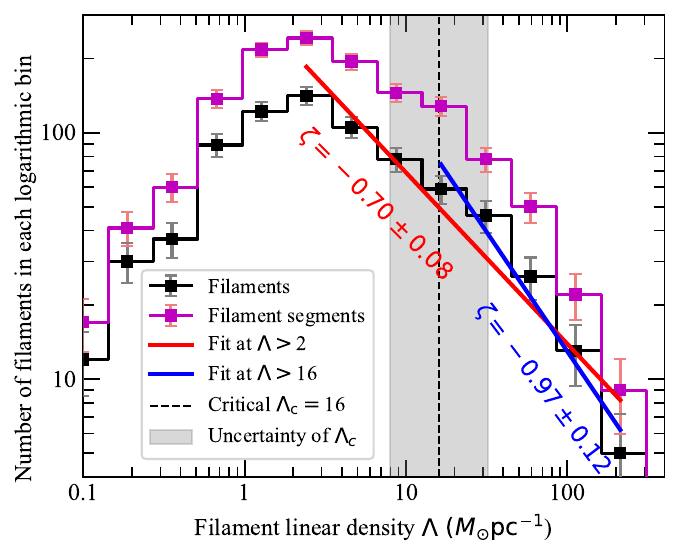}
\caption{Filament linear density function (FLDF) for the Ophiuchus molecular cloud. The distribution of linear densities of 769 measurable filaments for $\Lambda \ge 2\,M_{\sun}\,\text{pc}^{-1}$ follows a power law with $\zeta = -0.70 \pm 0.08$, whereas the FLDF of the denser filaments with $\Lambda \ge \Lambda_{\rm{c}} \approx 16\,M_{\odot}\,\rm{pc}^{-1}$ is somewhat steeper, with $\delta = -0.97 \pm 0.12$. The critical density $\Lambda_{\mathrm{c}}$ is indicated by the dashed vertical line and the presumed range (a factor of 2) of its uncertainties is shown by the gray shaded area. If the filaments are segmented into chunks of 0.1 pc length, the slope of the resulting FLDF remain almost identical. The turnover at $\Lambda_{0} \approx 2\,M_{\sun}\,\text{pc}^{-1}$ indicates that the extracted filaments may be incomplete in the low-density end.}
    \label{fig:FLDF}  
\end{figure}

\subsection{Linear Density Function of the Filaments}
\label{lindensfun}

Filamentary structures in molecular clouds are essential for understanding the star formation processes, because dense cores are usually found (and presumably formed) within dense filaments \citep{Andre+2014,Konyves+2015}. The filament linear density function (FLDF) is an important characteristic of the distribution of filaments over their linear densities $\Lambda$, similar to the core mass function \citep{Zhang+2024}. Defined analogously, it also shows a power-law distribution ${\rm d}N / {\rm d}\log\Lambda \propto{\Lambda}^{\zeta}$ at high linear densities. Analyses of \textit{Herschel} observations showed that $\zeta \approx -1.5$ for dense filaments with $\Lambda \ga \Lambda_{\rm c} \approx 16\,M_{\odot}\,{\rm pc}^{-1} $ \citep{Andre+2019a, Zhang+2024}. Filaments with such linear densities are expected to become gravitationally unstable and fragment into dense cores \citep[e.g.,][]{Zhang+2020}. However, the critical value $\Lambda_{\rm c} \approx 16\,M_{\odot}\,{\rm pc}^{-1}$ may be uncertain within a factor of $\sim 3$ \citep[e.g.,][]{Li+2023} and, therefore, it should be considered only as a rough indicator of a filament instability.

The linear densities of 769 measurable filaments were computed by \textit{getsf} as the ratio of the filament mass $M_{\rm F}$ to its length $L_{\rm F}$ \citep{Menshchikov2021method}. For more accurate results, we selected one-sided measurements of $\Lambda$ or arithmetic averages from both filament sides, following the approach we used to select the well-measurable widths of filaments (Sect.~\ref{filamentprops}). In other words, we adopted a good median width of a filament as a proxy to determine the goodness of linear density measurements. 

The FLDF for the Ophiuchus molecular cloud shows a shallow slope of $-0.70 \pm 0.08$ in the range of linear densities $2 < \Lambda < 300\,M_{\odot}\,\rm{pc}^{-1}$ (Fig.~\ref{fig:FLDF}). The densest, likely gravitationally unstable filaments with $\Lambda_{\mathrm{c}} < \Lambda < 300\,M_{\odot}\,\rm{pc}^{-1}$, display a somewhat steeper slope of $-0.97 \pm 0.12$. In principle, local physical conditions in filaments must be more relevant for the onset of instabilities and fragmentation of the filaments into cores than average properties of the entire (sometimes long) filaments. We explored this idea by producing another FLDF, based on short segments of the filaments. Following the approach used by \cite{Zhang+2024}, we segmented all filaments into 0.1 pc chunks, a scale of the typical half-maximum width of the observed filaments. As shown in Fig.~\ref{fig:FLDF}, the segmented filaments produced an almost identical shape with the slopes of $-0.70 \pm 0.08$ for $\Lambda > 2\,M_{\odot}\,\rm{pc}^{-1}$ and $-1.03 \pm 0.09$ for $\Lambda > \Lambda_{\mathrm{c}}$. The two approaches give, therefore, consistent results for the Ophiuchus cloud.

\section{Structural Analysis of the L1688 HFS}
\label{StructAnalysis}

\begin{figure}
     \centering
     \includegraphics[width=1.0\hsize]{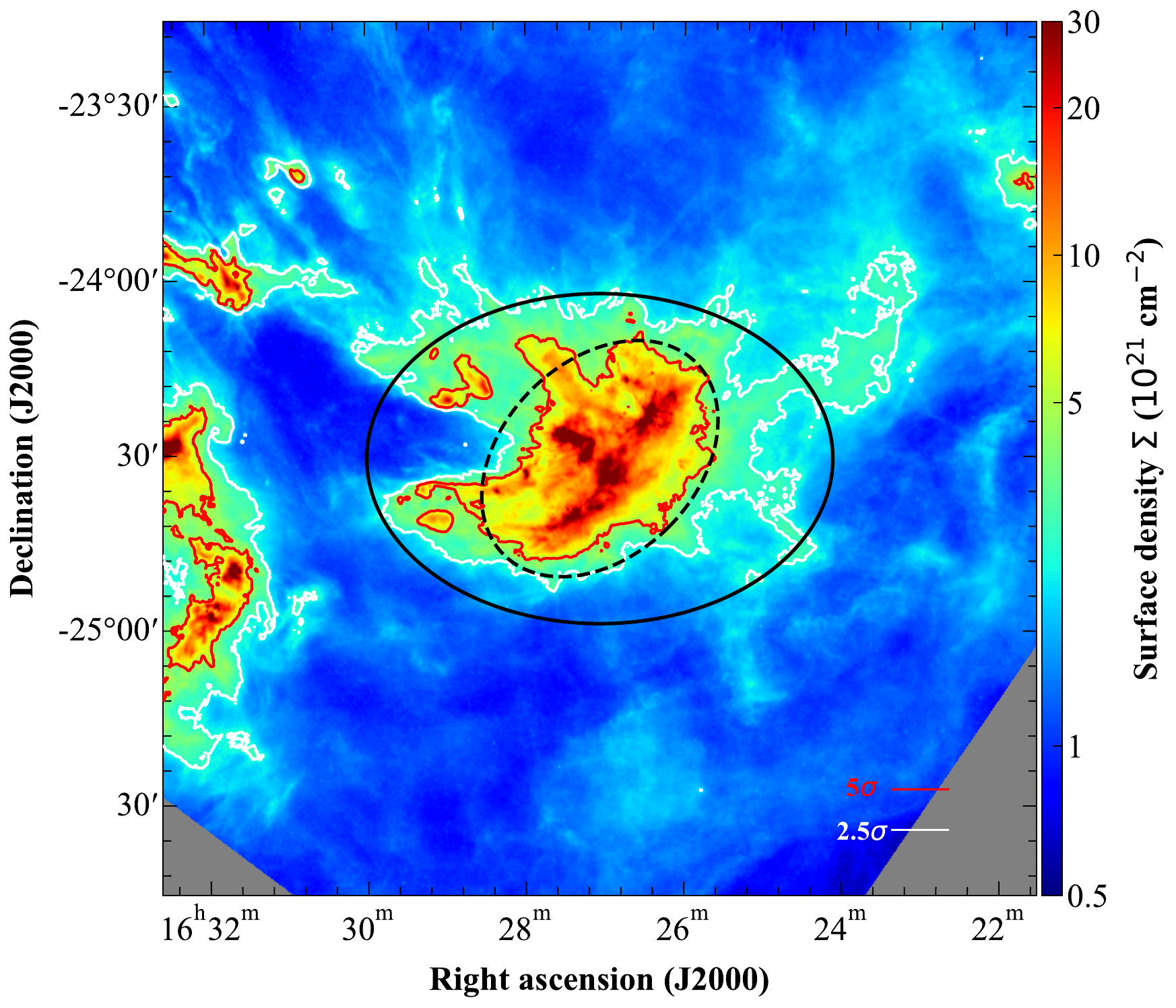}
     \caption{
Morphology and extent of the Ophiuchus L1688 hub, delineated by the ellipses encompassing the areas above 2.5\(\sigma\) and 5\(\sigma\) surface density fluctuations levels (white and red contours, respectively). The ellipses with semi-major and semi-minor axes of 2400 and 1700\arcsec (1.7 and 1.2 pc) and 1400 and 1000\arcsec (1.0 and 0.7 pc) include the entire hub region and its densest area, respectively.}
     \label{fig:1688S}  
\end{figure}

Hub-filament systems are the regions within molecular clouds, where multiple filaments converge, characterized by high surface densities and compact rounded morphologies \citep{Myers+2009,Schneider+2012,Peretto+2014,Chen+2019,Dib+2020,Kumar+2020,Xu+2023}. In star-forming regions, such as L1688 in the Ophiuchus molecular cloud, hubs are the networks of short, high-density filaments (Fig.~\ref{fig:1688C}) rather than single massive clumps \citep{Kumar+2020}. Hubs concentrate mass and serve as the sites for star formation, facilitating the coalescence of filaments and directing the flow of material that leads to the formation of dense cores \citep{Schneider+2012,Kumar+2020}.

\subsection{The Hub Morphology}

Figure~\ref{fig:1688S} delineates the hub shape and extent in the surface density map with two ellipses, defined at 2.5$\sigma$ and 5$\sigma$ fluctuations levels, where $\sigma = 10^{21}$ cm$^{-2}$ was estimated in the source- and filament-free regions. The ellipses have semi-major and semi-minor axes of 2400 and 1700\arcsec (1.7 and 1.2 pc) and 1400 and 1000\arcsec (1.0 and 0.7 pc) and they are centered at RA $=$ 16h 27m 04s and Dec $= -24^\circ$ 30{\arcmin} 45{\arcsec}. The smaller ellipse (position angle PA $= 135^\circ$) 
includes the inner dense area of the hub and the larger ellipse (PA $= 90^\circ$) encompasses the entire hub extent (cf. Fig.\ref{fig:1688Mline}). A cross-verification with the low-resolution Planck data contours confirmed the ellipse parameters. It is worth noting that the hub extent in the L1688 region is similar to that measured by \citet{Dib+2020} for the hubs in Cygnus-X North.

\begin{figure}
 \centering
 \includegraphics[width=1.0\hsize]{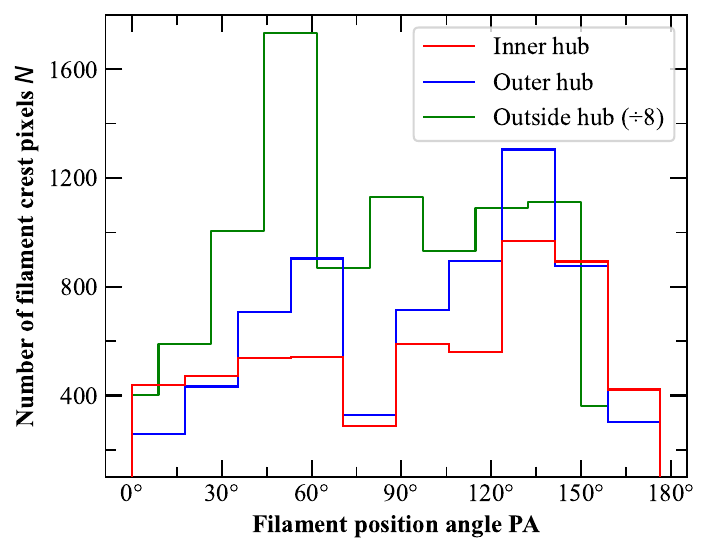}
 \caption{Distributions of filament orientations (position angles) for different areas in the Ophiuchus molecular cloud. The histograms for the inner and outer parts of the L1688 hub and outside area show the numbers of filament crest pixels across position angles. For presentation, the counts for the outside region were scaled down by a factor of 8.}
 \label{fig:1688Anglehisto}  
\end{figure}

\begin{figure*}
\centering
\includegraphics[height=0.31 \hsize]{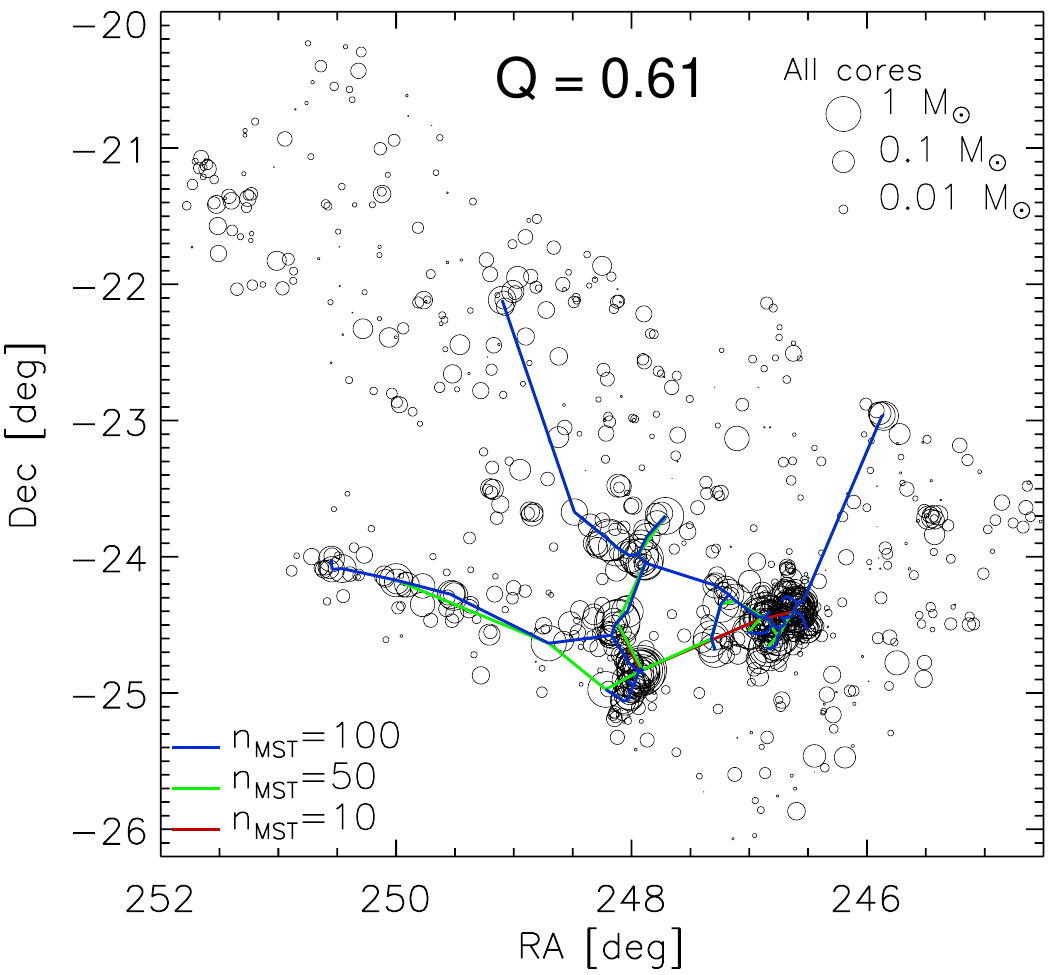}
\includegraphics[height=0.31 \hsize]{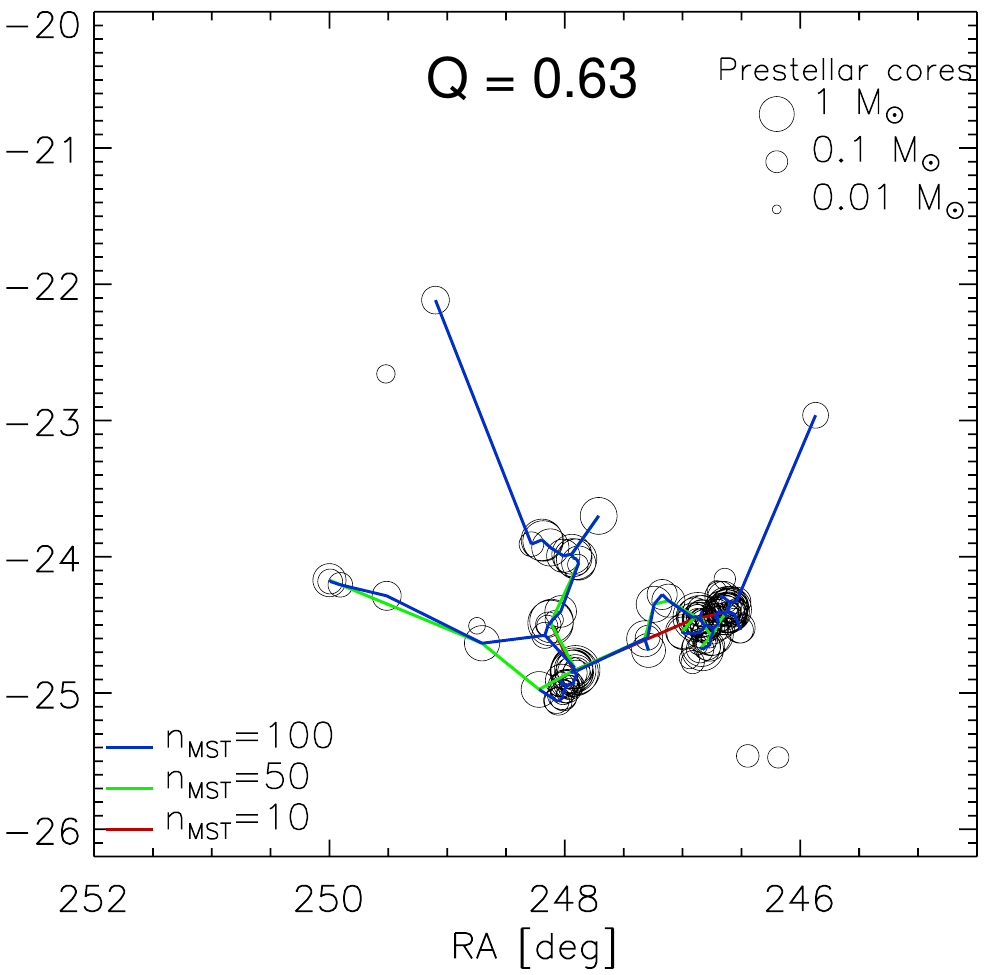}
\includegraphics[height=0.31 \hsize]{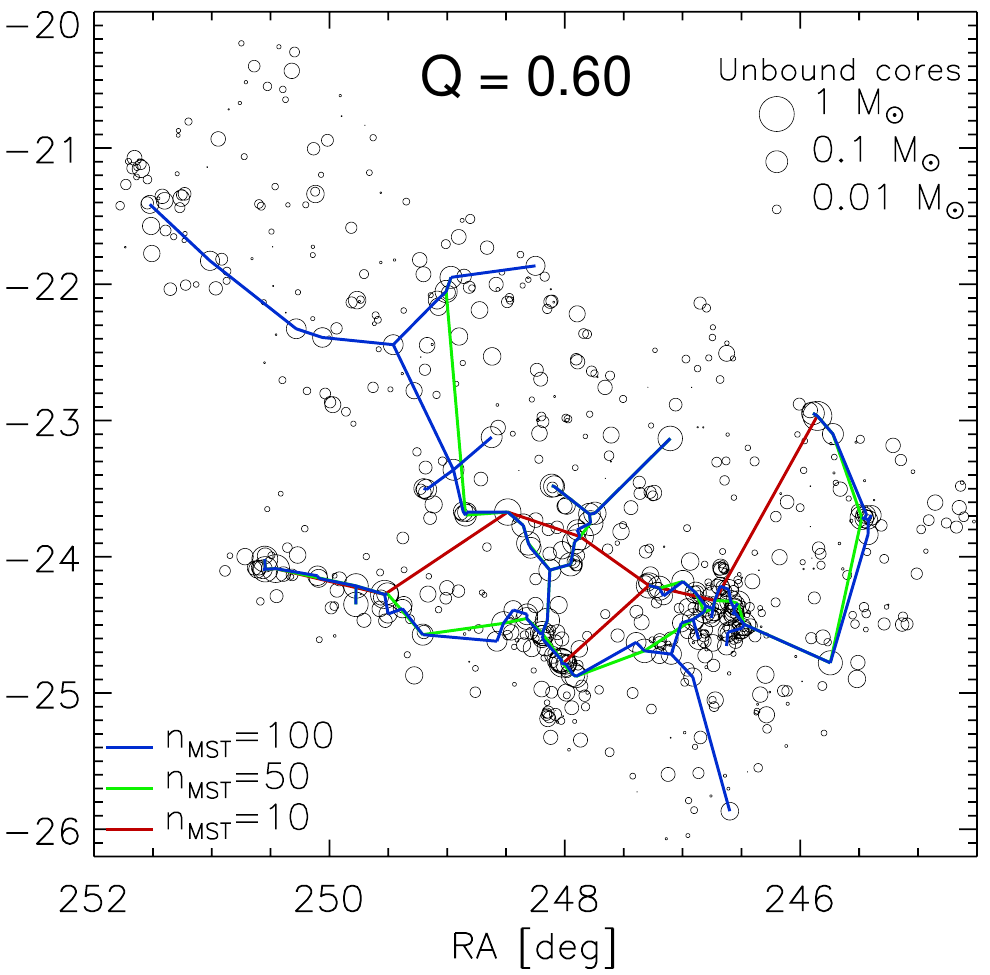}
\caption{MST results for the entire ensemble of starless cores (left panel), prestellar (candidate and robust) cores (middle panel) and unbound cores (right panel) in the Ophiuchus molecular cloud. Core sizes are arbitrarily scaled by their masses, highlighting the positions of the most massive cores. The MST for the 10, 50, and 100 most massive cores are shown with the red, green, and blue lines, respectively.} 
\label{fig:ophiuchus_core_distribution}
\end{figure*}

The L1688 hub encloses dense filamentary structures that are roughly parallel to the major axis of the inner ellipse (Fig.~\ref{fig:1688C}), as indicated by the position angles of the filamentary structures in the L1688 HFS. Two prominent peaks in the filament orientations within the hub are almost orthogonal to each other, at PA ${\approx\,}135^\circ$ and $50^\circ$ (Fig.~\ref{fig:1688Anglehisto}). A significant fraction of the filamentary structures outside the hub region is also aligned at PA ${\approx\,}50^\circ$. The orientations are consistent with previous studies that found non-random alignments of filaments in star-forming regions \citep[][]{Goldsmith+2008, Palmeirim+2013}.

These results suggest that the filamentary structures in the L1688 HFS exhibit a non-random, preferential alignment, particularly outside the central hub, which might indicate the influence of the hub’s gravitational potential or other localized processes affecting the alignment of filaments. We refer to \cite{Andre2017} for a discussion of the gravitational and magnetic effects on filament alignment.

\subsection{Spatial Distribution of Cores}
\label{SpatialDistr}

The structure parameter $\mathcal{Q}$ is a quantitative measure for assessing the spatial distribution of stars within clusters, particularly in distinguishing between the centrally condensed and fractal-like substructured configurations. It has been widely applied in the analysis of both young and evolved star clusters \citep{Gouliermis+2012, Fernandes+2012, Delgado+2013, Parker+2014, Gregoriohetem+2015, Dib+2018}, as well as in studies of the spatial distribution of dense cores and young stars in star-forming regions, such as Aquila, Taurus, Orion B, and W43 \citep{Guthermuth+2009, Alfaro+Romanzuniga2018, Parker2018, Dib+Henning2019}, and distant massive clumps \citep{Xu+2024}.

The parameter $\mathcal{Q}$ is defined as the ratio of the normalized mean edge length of the minimal spanning tree (MST) to the normalized correlation length of the star cluster: $\mathcal{Q} = {\bar{m}}/{\bar{s}}$, where $\bar{m}$ represents the mean edge length of the MST, normalized by ${(N_{\text{tot}} \,A)^{1/2}} (N_{\text{tot}} - 1)^{-1}$, with $A$ denoting the cluster area and $N_{\text{tot}}$ the total number of stars. The mean separation $\bar{s}$ between stars is normalized by the overall cluster radius \citep{Cartwright+Whitworth2004}. The $\mathcal{Q}$ parameter is particularly valuable in distinguishing cluster morphologies; values $\mathcal{Q} > 0.8$ are typically associated with centrally condensed clusters that exhibit a smooth radial density gradient following a power-law distribution $\rho \propto r^{-\alpha}$. In contrast, values $\mathcal{Q} < 0.8$ suggest a more hierarchical or fractal structure characterized by significant subclustering \citep{Cartwright+Whitworth2004,Schmeja+Klessen2006}.

Figure~\ref{fig:ophiuchus_core_distribution} shows the spatial distribution and MSTs of the 10, 50, and 100 most massive cores for three sets of extracted cores in the Ophiuchus molecular cloud. The overall spatial distribution of the entire set of 818 starless cores is substructured and fractal-like, as indicated by $\mathcal{Q} = 0.61$. When evaluated separately, the sets of 132 (candidate and robust) prestellar and 686 unbound cores are described by $\mathcal{Q} = 0.63$ and 0.60, respectively. The relatively higher $\mathcal{Q}$ value obtained for the prestellar cores suggests a slightly more evolved and centrally concentrated state, compared to a more dispersed distribution of the unbound cores. The unbound cores are more numerous, therefore they heavily influence the overall distribution, resulting in a close resemblance of the $\mathcal{Q}$ value for the entire sample to that of the unbound cores.

\subsection{Mass Segregation of Cores}
\label{MassSegr}

The parameter $\mathcal{Q}$ does not contain information about the core mass segregation. The mass segregation ratios $\Lambda_{\rm MSR}$ and $\Gamma_{\rm MSR}$ provide quantitative measures of how the massive cores are distributed relative to the lower-mass ones \citep{Allison+2009}. The mass segregation ratio is defined as $\Lambda_{\rm{MSR}} = \langle\mathcal{L}^{\rm{rand}}_{\rm{MST}}\rangle /\mathcal{L}^{\rm{mp}}_{\rm{MST}}$, where the nominator is the average MST length for a randomly selected subset of cores and the denominator is that for the most massive cores. Values $\Lambda_{\rm MSR} > 1$ indicate mass segregation, when the massive cores are more centrally concentrated.
Conversely, $\Lambda_{\rm MSR} < 1$ points to inverse mass segregation, when the massive cores are less centrally concentrated, whereas $\Lambda_{\rm MSR} \approx 1$ suggest a random distribution of massive cores. Some authors suggested, however, that mass segregation corresponds to $\Lambda_{\rm MSR} > 2$ \citep[e.g.,][]{Dib+Henning2019}.
The MST-based mass segregation description was refined by \citet{Olczak+2011} in a parameter $\Gamma_{\text{MSR}}$. The redefinition, which incorporates the use of a geometric mean as an intermediate step, enhanced the method sensitivity, enabling a more robust detection of lower levels of mass segregation. 

Mass segregation in star-forming regions and clusters was evaluated using these methods by several groups \citep{Parker2018,Dib+2018,Dib+Henning2019,Sadaghiani+2020,Paulson2024}. Notably, \citet{Dib+Henning2019} provided a comprehensive analysis of mass segregation in star-forming regions, particularly focusing on the correlation between the structure of molecular clouds and their star formation activity. They found that regions of star formation with higher surface densities (such as W43) exhibit higher levels of mass segregation, with massive cores being more centrally concentrated. In contrast, regions like Taurus with a low star formation activity show no significant mass segregation, evidenced by $\Lambda_{\rm MSR}$ and $\Gamma_{\rm MSR}$ close to unity. 

\begin{figure}
\centering
\includegraphics[width=1.0\hsize]{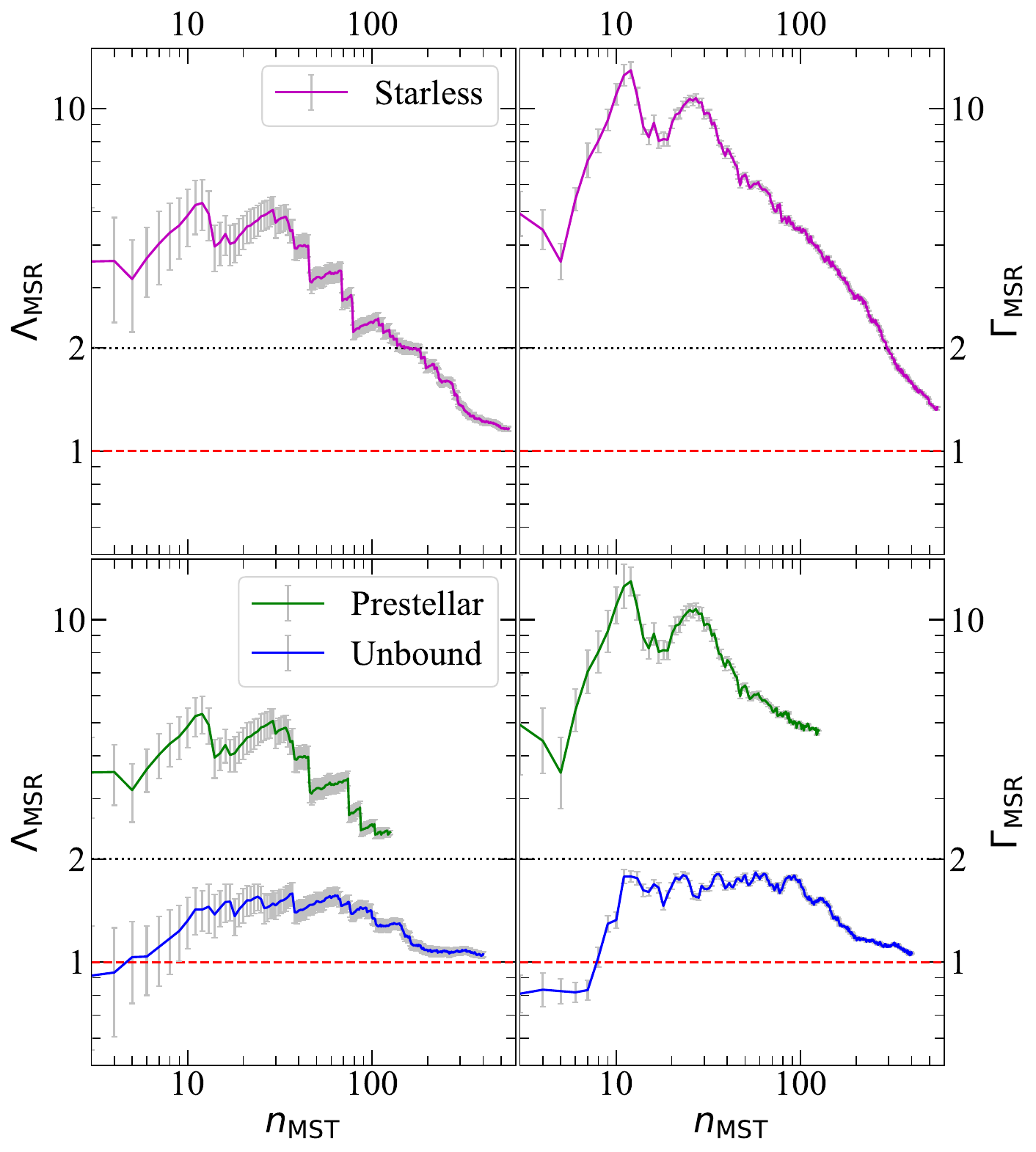}
\caption{Mass segregation ratios $\Lambda_{\text{MSR}}$ (left panels) and $\Gamma_{\text{MSR}}$ (right panels) for the entire set of 818 starless cores, 132 (candidate and robust) prestellar cores, and 686 unbound cores in the Ophiuchus molecular cloud as functions of the numbers $n_{\text{MST}}$ of the most massive cores. The error bars are the standard deviations derived from 100 realizations for each sample size. The dashed and dotted horizontal lines bound the range of values that likely correspond to non-segregated distributions of cores.}
\label{fig:msr_massive_cores}
\end{figure}

Figure~\ref{fig:msr_massive_cores} displays the mass segregation ratios for the entire set of 818 starless cores, 132 (candidate and robust) prestellar cores, and 686 unbound cores in the Ophiuchus molecular cloud. We calculated $\langle\mathcal{L}^{\rm{rand}}_{\rm{MST}}\rangle$ for both prestellar and unbound cores by selecting 100 random sets of $n_{\rm MST}$ cores from the 818 starless cores. The same sample for the random selection facilitates comparison of the segregation values for the sub-samples of cores. The results for the entire set of starless cores show very significant mass segregation ($\Lambda_{\rm{MSR}} \approx$ 4--5 and $\Gamma_{\rm{MSR}} \approx$ 5--10) for up to $n_{\rm{MST}}\approx 40$. The mass segregation of larger numbers of the most massive cores continuously declines until it essentially vanishes for $n_{\rm{MST}}\ga 300$ ($2 \gtrsim \Lambda_{\rm{MSR}}, \Gamma_{\rm{MSR}} \rightarrow 1$). The sample of prestellar (candidate and robust) cores (Fig.~\ref{fig:msr_massive_cores}) is mass-segregated almost identically to the entire set of extracted cores for $n_{\rm{MST}}\la 100$. The strongest mass segregation is exhibited by the  $\sim 60$ most massive prestellar cores with $M > 0.5$ $M_{\sun}$ (Fig.~\ref{fig:CMF}), indicating that the cores are spatially clustered. In contrast, the sample of unbound starless cores shows a more dispersed spatial distribution for $n_{\rm{MST}}\la 100$ with very low or no mass segregation ($\Lambda_{\rm MSR} \la 1.5$ and $\Gamma_{\rm MSR} \la 2$). Therefore, mass segregation is the property that markedly separates the (more massive) prestellar cores from the unbound cores.

\begin{figure*}
   \centering
   \includegraphics[width=1.0 \hsize]{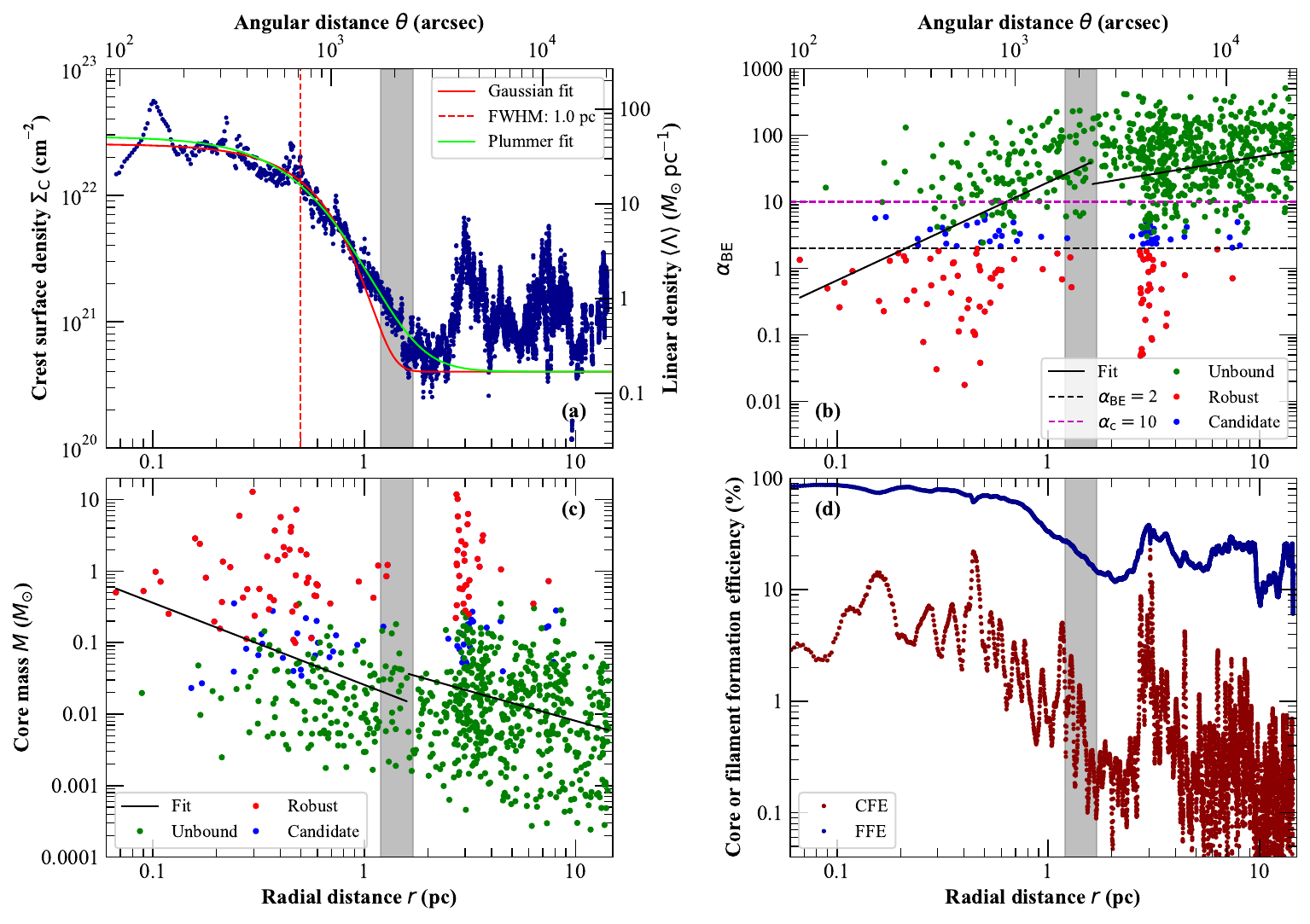}
    \caption{Properties of filaments and cores in the Ophiuchus molecular cloud as a function of the distance $r$ from the L1688 hub center. The gray shaded area indicates the position of the elliptical hub boundary at $1.2 \le r \le 1.7$ pc. {\bf (a)} Crest surface densities $\Sigma_{\rm C}$ of all 769 measurable filaments, averaged over circular annuli of 1 pixel width, and the corresponding linear densities $\langle \Lambda \rangle$ for $\langle W \rangle = 0.12$ pc. The red and green lines show a Gaussian profle with FWHM $= 0.99 \pm 0.02$ pc and a Plummer profile $\Sigma_{\rm C}(r) = \Sigma_{0}\,(1 + (r / R_{0})^2)^{-2.5}$ with $\Sigma_{0} = (2.84 \pm 0.34) \times 10^{22}$ $\mathrm{cm}^{-2}$ and $R_{0} = 0.76 \pm 0.02$ pc, respectively, with an additional background value of $4 \times 10^{20}\, \mathrm{cm}^{-2}$. {\bf (b)} Bonnor-Ebert stability parameter $\alpha_{\rm BE}$ for all 818 starless cores of the three types, fitted by $\alpha_{\rm BE}(r) = 19.5\,r^{1.48}$ inside the hub and $\alpha_{\rm BE}(r) = 14.5\,r^{0.53}$ outside it. The dashed horizontal lines show the limiting values $\alpha_{\rm BE} = 2$ and $\alpha_{\rm c} = 10$ (unresolved case, with $O/H = 1$), used to select the 85 robust and 47 candidate prestellar cores (Sect.~\ref{classification}). {\bf (c)} Masses $M$ of the starless cores of the three types, fitted by $M(r) = 0.026\,r^{-1.15}$ in the hub and $M(r) = 0.55\,r^{-0.83}$ outside the hub. {\bf (d)} Core and filament formation efficiencies within circular annuli of 1 pixel width as functions of $r$. Estimated more globally, the respective CFE and FFE values are $4.9$\% and $71$\% inside the dense hub, $1.3$\% and $31$\% in the outer zone of the hub, and $0.86$\% and $21$\% outside the hub (Fig.~\ref{fig:1688S}).
    }
\label{fig:1688Mline}
\end{figure*}

\subsection{Radial Distribution of Filament and Core Properties}

The properties of filaments and cores significantly depend on the distance from the L1688 hub center. Figure~\ref{fig:1688Mline} shows the radial dependence of the filament crest surface densities $\Sigma_{\rm C}$, average linear densities $\langle \Lambda\rangle$, and the stability parameter $\alpha_{\rm BE}$ for the starless cores. The $\Sigma_{\rm C}$ values of all 769 measurable filaments were averaged within concentric rings of 1 pixel width, originating at the hub center, whereas the $\alpha_{\rm BE}$ values correspond to each of the selected 818 starless cores. 

With increasing distance, both $\Sigma_{\rm C}$ and $\langle \Lambda\rangle$ decrease almost by two orders of magnitude (Fig.~\ref{fig:1688Mline}), revealing a well-developed Gaussian-like (Plummer) profile $\Sigma_{\rm C}(r)$ of the hub and a transition from its dense structures to the less denser filaments in the molecular cloud outside the hub ($r > 1$ pc). There are several clear, relatively strong peaks in $\Sigma_{\rm C}(r)$ outside the hub (at $r \approx 3$, 7, and 18 pc), that correspond to the dense parts of the filamentary structures in the Ophiuchus molecular cloud (Fig.~\ref{fig:1688C}). 

The gravitational stability parameter $\alpha_{\rm BE}(r)$ reveals that most of the bound starless cores tend to reside within the hub area (Fig.~\ref{fig:1688Mline}). Although there is a large scatter in $\alpha_{\rm BE}$, on average the values tend to significantly increase toward larger distances from the center, indicating that the starless cores tend to become progressively less bound within the hub and mostly unbound outside it. Distribution of the masses $M(r)$ of starless cores within the hub is (approximately) inversely proportional to $r$ (Fig.~\ref{fig:1688Mline}), although there is a large scatter in the masses. Outside the hub, the core masses remain much scattered, but tend to follow a shallower average distribution. The population of candidate and robust prestellar cores is mostly located inside the hub and at two radial locations outside it ($r \approx 3$ and 7 pc). The fact that the two radii coincide with the peaks in $\Sigma_{\rm C}(r)$, indicates that the prestellar cores belong to the dense filaments found in the molecular cloud at these distances from the hub (Fig.~\ref{fig:1688C}).

\subsection{Core and Filament Formation Efficiencies}

The \textit{Herschel} images and surface density map were decomposed by \textit{getsf} in separate images of the structural components of sources, filaments, and their backgrounds \citep{Menshchikov2021method}. Based on the component separation, we analyzed the radial mass distribution of the structural components, examining the core formation efficiency (CFE) and filament formation efficiency (FFE) as functions of the radial distance from the hub center. The formation efficiencies are defined as the mass ratios of the dense cores and filaments, respectively, to the total mass of the molecular cloud \citep{Zhang+2018}. 

Our results show that both CFE and FFE in the Ophiuchus molecular cloud vary significantly with the radial distance $r$ from the hub center (Fig.~\ref{fig:1688Mline}). The CFE values display much larger fluctuations than the FFE values do, because of the much greater spread of the core masses $M$ than of the surface densities $\Sigma_{\rm C}$ within the circular annuli (Fig.~\ref{fig:1688Mline}). On average, however, CFE and FFE demonstrate quite similar behavior, decreasing from the maximum values at the hub center toward much lower values at the hub boundary and in the cloud outside the hub, where they exhibit several local peaks. This is likely due to the elongated and asymmetric distribution of filaments around the hub, which causes locally enhanced core formation along certain directions. It should be noted that a particularly prominent peak at a radial distance of $\sim$3\,pc corresponds to the eastern and southeastern parts of the cloud, where the nearby star-forming regions L1709 and L1689 are located. These regions host multiple filaments and dense cores, which contribute significantly to the elevated CFE and FFE at that distance. Logarithmic values of FFE and CFE are positively correlated within the hub (Pearson correlation coefficient $p = 0.78$), whereas the correlation becomes weaker outside the hub ($p = 0.52$). The significant correlation suggests that the filamentary hub may be enhancing the efficiency of star-formation processes.

\section{Discussion}
\label{discuss}

\subsection{CMF, FLDF, and Their Implications for Star Formation}

Empirical and numerical studies of the CMF consistently show that the number of cores declines with increasing mass, particularly at the higher-mass end of the mass distribution \citep{Klessen+2005,Dib+2008,Andre+2010,Anathpindika2013}. A power-law behavior is thought to result primarily from the self-similar and hierarchical nature of turbulence and fragmentation in molecular clouds \citep{Larson1981, Elmegreen+Falgarone1996, Padoan+Nordlund2002, Federrath+Klessen2012, Myers+2014}. The CMF is often seen as a precursor to the IMF, suggesting that the stellar mass distribution is inherited from the mass distribution of prestellar cores \citep{Alves+2007, Zhang+2024}. 

Our analysis (Sect.~\ref{coremassfun}) reveals relatively shallow slopes of the CMF of prestellar cores ($\delta \approx -0.53$ to $-0.86$) in the Ophiuchus molecular cloud, compared to the IMF ($\delta = -1.35$), consistently with several other observational studies of star-forming regions \citep{Li+2007, Zhang+2015, Marsh+2016, Zhang+2018, Pouteau+2022}. The shallow slopes might suggest that the relationship between CMF and IMF depends on some additional factors, such as the environments and evolution of the cores. 
Gas accretion by the cores can play a pivotal role in the transition from the CMF to the IMF by allowing the cores that are near the critical mass for collapse to grow and eventually form stars \citep{McKee+Ostriker2007, Hennebelle+Chabrier2008, Dib+2010b}. In the densest environments, collisions between the cores can lead to the formation of more massive cores and to modifications of the CMF shape inherited from turbulent fragmentation \citep{Dib+2007b,Dib+2023}. Additionally, other theoretical and numerical studies have demonstrated that feedbacks from stellar winds, radiation, and outflows can significantly alter core growth by restricting accretion and potentially dispersing the low-mass cores \citep{Krumholz+McKee2005, Dale+2005, Padoan+2017}.

However, the CMFs based on the masses derived by SED fitting may not be accurate enough (Sect.~\ref{coremassfun}) to draw reliable conclusions on star formation. Errors and biases of the derived masses \citep{Men2016fitfluxes} can redistribute the cores between the mass bins with respect to the true CMFs of the physical objects and alter the intrinsic shape of the mass functions. 
Nevertheless, if the adopted mass bins are larger than or comparable to the typical uncertainties in mass, the overall shape of the CMF remains relatively robust, as individual cores are unlikely to move across multiple bins due to uncertainties alone.
Simulated images may not be fully consistent with the observed images, therefore the limiting mass obtained in core extraction completeness simulations may be inaccurate. 
For these reasons, while the CMF provides useful insights into the core population, we stress the need to be cautious in its astrophysical interpretations and to keep in mind that the observationally derived mass functions may still carry significant implicit uncertainties.

Our MST analysis (Sects.~\ref{SpatialDistr} and \ref{MassSegr}) reveals a clustered distribution of massive cores in the Ophiuchus molecular cloud, indicating a higher degree of central concentration driven primarily by core mass. This central concentration enhances the gravitational potential, facilitating further accretion by massive cores and reinforcing the link between the CMF and IMF. While some massive cores remain unbound, they are expected to become bound within $10^3$ to $10^4$ years through continued accretion \citep[e.g.,][]{Zhang+Tan2011,Zhang+2023}. Regions with deeper gravitational potentials, associated with higher star-formation rates, facilitate this process, consistent with observations in other star-forming regions \citep{Parker+Goodwin2015, Dib+Henning2019}. However, the MST also identifies a significant fraction of spatially dispersed, low-mass unbound cores that are unlikely to accumulate enough mass to become bound, consistent with the findings that many low-mass cores do not evolve into stars \citep{Padoan+Nordlund2002, WardThompson+2007, Francesco+2007, Offner+2014}. 

FLDFs offer an additional perspective on mass distribution along filaments in molecular clouds, complementing insights from the CMFs \citep{Andre+2019a, Zhang+2024}. For the Ophiuchus molecular cloud, we found the FLDF with the power-law slopes $\zeta = -0.70$ to $-0.97$ (Sect.~\ref{lindensfun}), similar to those of the CMF and shallower than the Salpeter slope. Hierarchical filamentary structures, shaped by turbulent processes, can lead to a broad range of filament linear densities $\Lambda$ with a power-law distribution \citep{Padoan+Nordlund2002, Hennebelle+Chabrier2008}. 
Filaments with supercritical $\Lambda > \Lambda_{\rm c}$ are more prone to gravitational fragmentation into dense cores. It is conceivable that the shape of the resulting CMF could be consistent with the distribution of $\Lambda$ (masses per unit length) of their parent filaments. This would be in line with the central role of the gravitational instability in the formation of cores \citep{Inutsuka+Miyama1997, Toci+Galli2015, Andre+2014, Zhang+2020}. Other star-forming regions, such as the California molecular cloud, also show strong correlations between $\Lambda$ and core formation, with the denser filaments preferentially forming more massive cores \citep{Zhang+2024}. The shallower slopes of both the FLDF and CMF in Ophiuchus reinforce the idea that the filament properties significantly impact core mass distribution \citep{Andre+2010, Roy+2015}. Limited sensitivity and angular resolution of observations and the structural complexity of the dense molecular clouds affect the observed slopes by under-representing the lower-density filaments and lower-mass cores and underestimating their masses \citep{Andre+2014, Arzoumanian+2019, Menshchikov2023}. 

Our results emphasize significant roles of the core mass segregation, gravitational potential, and local environments in determining the evolution of starless cores and the eventual stellar population. CMFs and FLDFs are useful tools, but their predictive power is limited, particularly for the low-density filaments and low-mass cores, affected by incomplete sampling. For a more comprehensive understanding of star formation we need to take into account both the environmental factors and physical properties and processes governing the dynamical evolution (e.g., turbulence, gas pressure, temperature gradients, and gravitational collapse) that would determine, whether the filaments would eventually fragment into cores and the latter would ultimately form stars.

\subsection{Filament-Driven Core Formation in the L1688 HFS}   

The filament alignment within the L1688 HFS reveals two distinct position angle peaks at PA $\approx 50^\circ$ and $135^\circ$, suggesting that external forces influence their orientation. The primary component at PA $\approx 135^\circ$ aligns with the orientation of the main axis of the hub (Fig.~\ref{fig:1688S}). The PA $\approx 50^\circ$ component is suggestive of a directed material flow roughly from the north-east side, because the opposite south-west side displays almost no alignment of filaments in that direction (Fig.~\ref{fig:1688C}). Such an asymmetry could be created by the influence of large-scale external forces, likely from the nearby Sco OB2 association, located at a distance of $11 \pm 3$ pc from the Ophiuchus molecular cloud. 

The Sco OB2 association exerts feedback pressure on L1688, compressing the molecular cloud and facilitating the formation of dense filamentary structures \citep{Howard+2021}. The region between the B stars S1 and HD\,147889 shows signs of localized heating and compression, which supports the idea that external feedback has its role in shaping the filaments \citep[e.g.,][]{Abergel+1996, Liseau+1999, Wilking+2008}. Such feedback could enhance star formation by enhancing the density of filaments and material accretion into the hub, contributing to the observed high star formation efficiency \citep{Schneider+2010, Peretto+2013}. The external pressure from Sco OB2 likely plays a critical role in determining both the orientation and mass distribution within the filaments, significantly impacting star formation processes in L1688 \citep{Loren+1986, Abergel+1996, Motte+1998, Liseau+1999, Johnstone+2000, Nutter+2006}.

Besides the external feedback from Sco OB2, effects of the magnetic fields and gas accretion on the filament orientations must also be considered. The magnetic fields are known to guide the flow of gas along filaments, where their alignment is controlled by the interaction between magnetic tension and gravitational forces. In star-forming regions, the magnetic fields can align either parallel or perpendicular to filaments, depending on the local density, influencing filament orientation \citep{Palmeirim+2013, Andre+2019b}. Studies like \citet{Planck+2016} show that the magnetic field orientations vary significantly between molecular clouds, shaping their filamentary structures. Gas accretion in the radial direction toward dense filamentary structures can increase the gravitational potential of hubs and align filaments toward the dense regions of star formation. Observations from systems like Taurus B211/B213 and California supercritical filaments suggest that such accretion is ongoing, with filaments acting as conduits for material flow \citep{Palmeirim+2013,Shimajiri+2019a,Zhang+2020}. Theoretical models support these ideas, with simulations of molecular cloud collapse showing that accretion onto filaments aligns with the observed mass inflow rates in star-forming environments \citep{Gomez+Vazquez2014,Vazquez+2019}.

Our analysis of the radial distribution of filaments indicates that the filaments inside the L1688 hub are dense and tightly packed, likely because of the gravitational forces, pulling the cloud material inward. With increasing distance from the hub center, the filament density decreases and they take a more diffuse configuration. This pattern is consistent with observations in other HFS regions, such as Serpens and Mon R2 \citep{Kirk+2013,Kumar+2022}. The positive correlation between FFE and CFE underscores the critical role of filaments in organizing mass and facilitating core formation. Regions, dominated by diffuse gas exhibit lower CFE and FFE, indicating the importance of dense filaments for core collapse and efficient star formation. The high CFE within the hub (reaching 5\%) demonstrates the efficiency of such filamentary environments in the production of prestellar cores. The significant decline of CFE outside the hub may support the hierarchical star formation model, where filaments act as the mass reservoirs that feed star-forming regions \citep{Schneider+2012,Chen+2019,Kumar+2020,Ren+2021,Ren+2023}. Our results suggest that the formation of prestellar cores in the L1688 HFS is predominantly driven by the filamentary structures that efficiently channel material into the hub.

\section{Conclusions}
\label{conc}

This study used the \textit{getsf} extraction method to analyze the \textit{Herschel} observations of the Ophiuchus molecular cloud, with a focus on the L1688 hub-filament system (HFS). By examining the structural and physical properties of the extracted filaments and cores, we derived the following results.

A total of 882 candidate cores were identified, including 85 robust prestellar cores, 47 candidate prestellar cores, 686 unbound starless cores, and 64 protostellar cores. A substantial fraction of the low-mass unbound cores (78\%) suggests that they will likely dissipate or merge together, rather than form stars individually. The core mass function (CMF) of the starless cores follows a power-law distribution with a relatively shallow slope of $\delta = -0.53$ over the masses $M$ of $0.04{-}10\,M_{\sun}$, compared to the Salpeter initial mass function (IMF) with $\delta =-1.35$. Although the most massive prestellar cores with $M >$ 1\,$M_{\sun}$ display a steeper power law with $\delta = -0.86$, the latter is still significantly shallower than the IMF slope.

Spatial distribution of the starless cores in the Ophiuchus molecular cloud indicates substructured, fractal-like configurations ($\mathcal{Q} = 0.60{-}0.63$). Mass segregation is prominent among the most massive cores, with only slight differences between the gravitationally bound and unbound cores. The low-mass unbound cores significantly influence the overall spatial distribution. Central clustering of the massive cores enhances the gravitational potential and promotes accretion in high-density regions, such as the L1688 HFS.

We identified 769 well-resolved filaments that have measurable widths, with a median half-maximum value $\langle W \rangle = 0.12$ pc, and 443 filaments that have measurable slopes of their profiles, with a median value $\langle \gamma \rangle = -1.4$ that corresponds to a power-law exponent $\langle \eta \rangle = -2.4$ of the volume density profiles. On average, the filament widths tend to increase by a factor of 4 with crest surface densities in the range of $10^{20}{-}10^{23}$ cm$^{-2}$, whereas the slopes show almost no average trends with the surface densities, although there is a large scatter in both quantities. 

The filament linear density function (FLDF) of the filaments reveals a power-law shape with a relatively shallow slope of $\zeta = -0.70$ over the linear densities $\Lambda$ of $2{-}300\, M_{\sun}\,{\rm pc}^{-1}$, consistent with the CMF slope that we find for the starless cores. The dense filaments with $\Lambda > \Lambda_{\rm c} \approx 16\, M_{\sun}\,{\rm pc}^{-1}$ display a steeper power law with $\zeta = -0.97$, also similar to the CMF slope for the massive prestellar cores with $M >$ 1\,$M_{\sun}$.

The filament and core formation efficiencies (FFE, CFE) in the Ophiuchus molecular cloud strongly depend on the radial distance from the hub center and are positively correlated. The CFE reaches high values of 5\% within the dense hub ($r \la 0.85$ pc), whereas it decreases to 0.9\% in the molecular cloud outside the hub. The FFE is as high as 71\% within the dense hub and it decreases to 21\% outside the hub, reflecting a transition from the filament-dominated hub to the background-dominated cloud at larger distances. 

Filaments seem to play a central role in concentrating mass and driving the core formation, particularly within dense star-forming hubs. They accrete gas, thereby promoting their gravitational fragmentation and the subsequent clustering of cores. An important problem is to study the feedback effects and how they affect the core and filament evolution. Expanding the sample size for various environments and incorporating regions at different distances are important for testing the general validity of our results and conclusions.

\begin{acknowledgements}
This work was supported by the National Natural Science Foundation of China (No. 11988101). DL is a New Cornerstone investigator. G.-Y. Z acknowledges additional support from the China Postdoctoral Science Foundation (No. 2021T140672), and the National Natural Science Foundation of China (No. U2031118). K.W. acknowledges support from the National Natural Science Foundation of China (12041305, 12033005), the China-Chile Joint Research Fund (CCJRF No. 2211), and the Tianchi Talent Program of Xinjiang Uygur Autonomous Region.

This work was also supported by the Key Project of International Cooperation of the Ministry of Science and Technology of China under grant number 2010DFA02710, as well as by the National Natural Science Foundation of China under grants 11503035, 11573036, 11373009, 11433008, 11403040, and 11403041. This work was carried out at the China-Argentina Cooperation Station of NAOC/CAS. 

We acknowledge contributions of the \textit{Herschel} Gould Belt Survey (HGBS) consortium and all researchers involved in the observations and data processing of the Ophiuchus molecular cloud, which laid down the foundation of this work. This research utilized data from the HGBS project, a Herschel Key Program conducted by the SPIRE Specialist Astronomy Group 3 (SAG 3), along with scientists from the PACS Consortium (CEA Saclay, INAF-IFSI Rome, INAF-Arcetri, KU Leuven, MPIA Heidelberg) and the Herschel Science Centre (HSC).
\end{acknowledgements}



\begin{appendix}

\section{High-resolution surface densities}
\label{sec:highres_density_maps}

Surface density maps are important for understanding the structure and physical properties of the cold interstellar medium. Derivation of the maps involves estimating zero offsets of the \textit{Herschel} images of a certain region by comparing them with the \textit{Planck} images of the same region \citep[e.g.,][]{Bernard+2010,Bracco+2020}. High-resolution surface density and temperature maps are then computed by fitting the spectral shapes of pixel intensities using the \textit{hires} method \citep{Menshchikov2021method}.

\begin{figure}[h]
\centering
\includegraphics[width=1.0 \hsize]{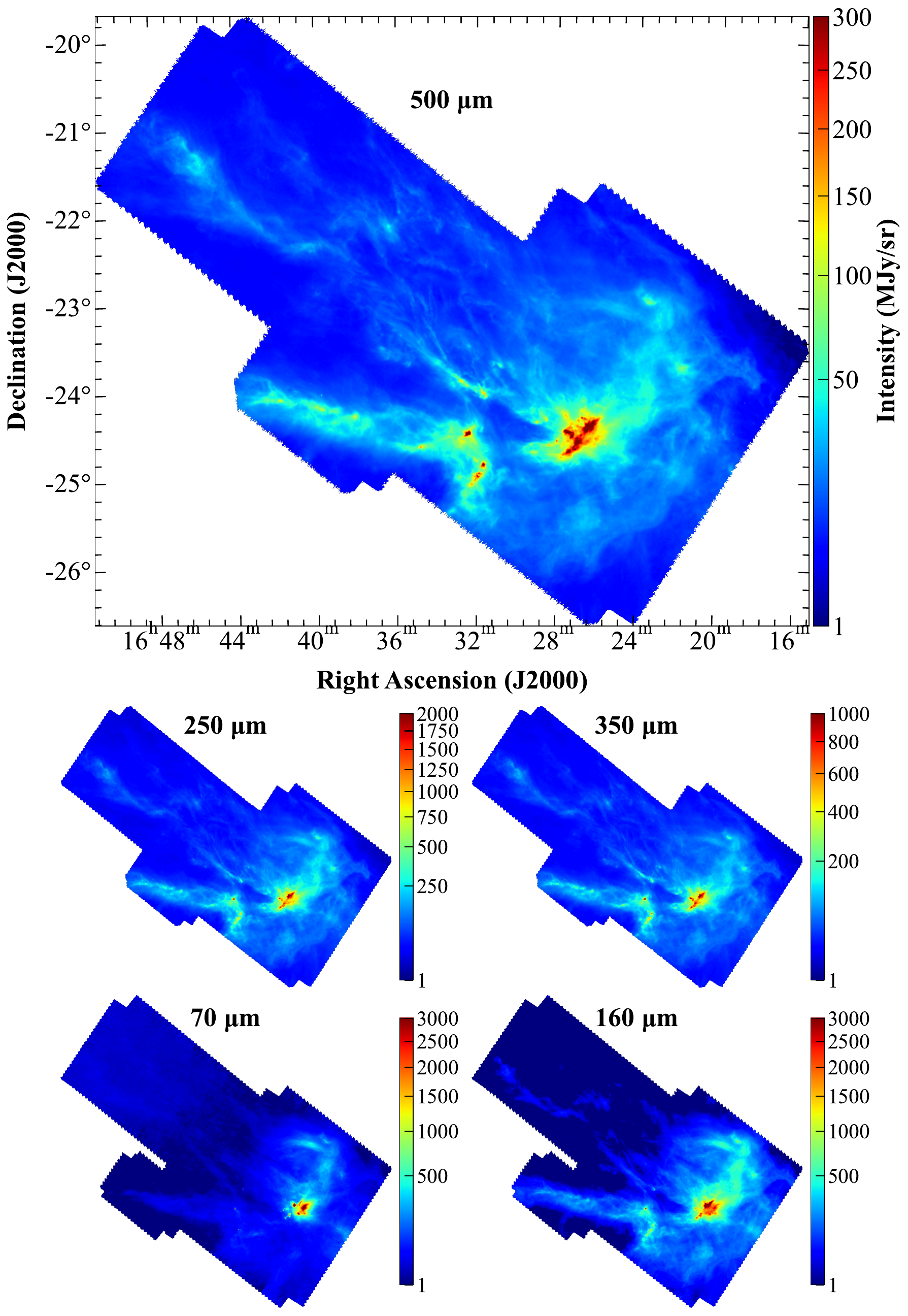}
\caption{\textit{Herschel} images of the Ophiuchus molecular cloud at 70, 160, 250, 350, and 500 $\mu$m and angular resolutions of 8.4, 13.5, 18.2, 24.9, and 36.3{\arcsec}, respectively.}
\label{Herscheloph}
\end{figure} 

\subsection{Zero Offsets for \textit{Herschel} Images and Derivation of Surface Densities and Temperatures}
\label{sec:offsets_surface_densities}
  
\textit{Herschel} imaged the Ophiuchus molecular cloud within its large 3.5 m aperture, in the wavelength range from 70 to 500 $\mu$m (see Fig.~\ref{Herscheloph}). However, the observational technique used by the \textit{Herschel} instruments could not guarantee accuracy of the intensities at the largest spatial scales. This deficiency could be reasonably well corrected using the so-called \textit{Planck} offsets \citep{Poglitsch+2010,Griffin+2010}. \textit{Planck} conducted an all-sky unbiased survey within its relatively small 1.5 m aperture, providing the images with a much lower, 5{\arcmin} angular resolution. The data, officially released by \textit{Planck}, are well-calibrated \citep{Planck+2014}, hence they can serve as a reliable standard for correcting the \textit{Herschel} images at their largest spatial scales.

\begin{figure}
\centering
\includegraphics[width=1.0 \hsize]{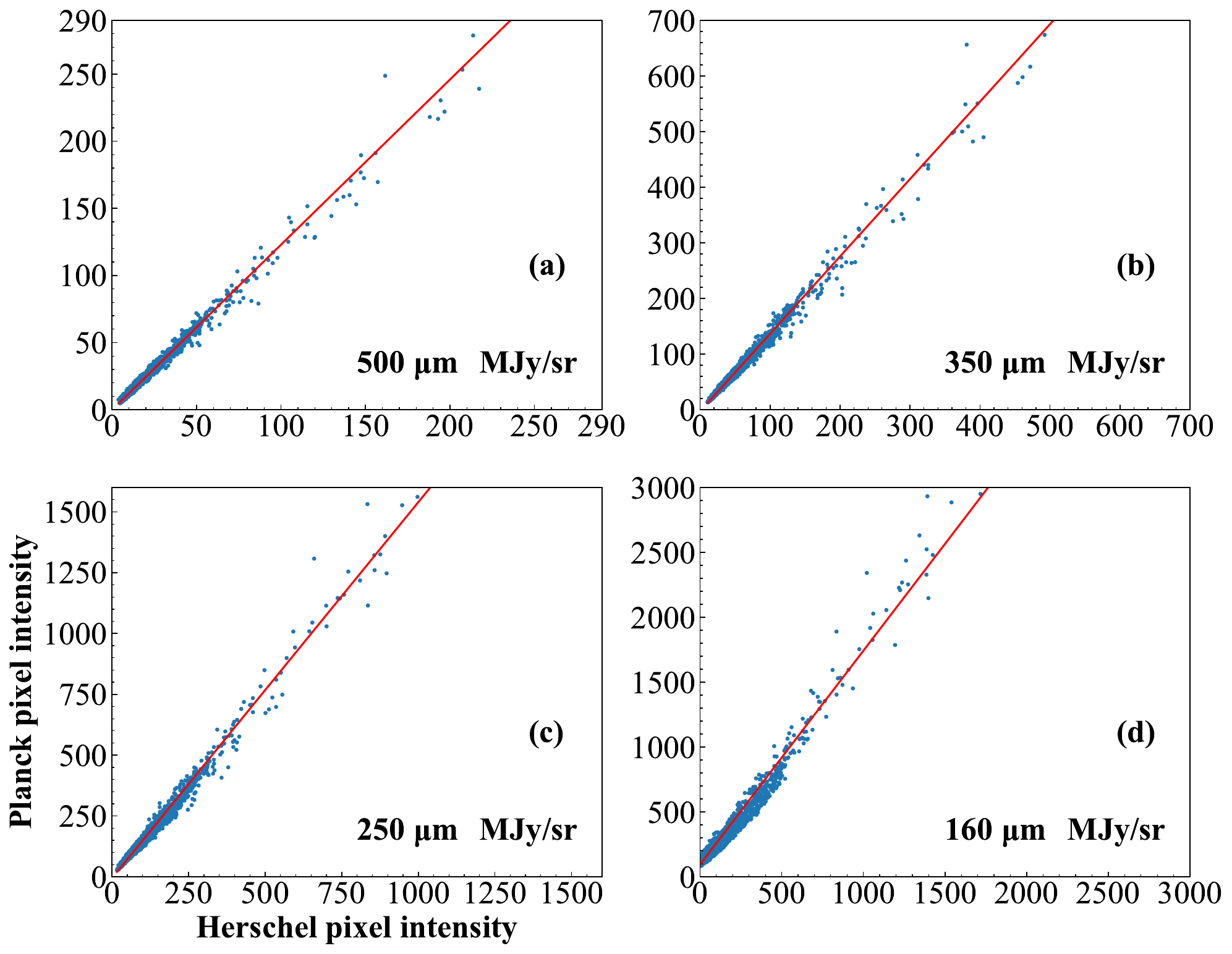}
\caption{Comparison of the pixel intensities in the \textit{Herschel} and \textit{Planck} images at the wavelengths 160--500 $\mu$m (the 70 $\mu$m image is not used). The red solid lines represent the linear fits and the blue dots correspond to the intensities of each pixel in the \textit{Herschel} and \textit{Planck} images. 
Median differences between the intensities are 134.2, 37.2, 12.8, and 3.5 MJy\,sr$^{-1}$ at 160, 250, 350, and 500 $\mu$m, respectively.}
\label{linfitoffsets}
\end{figure}

For the purpose of deriving the \textit{Planck} offsets for the \textit{Herschel} images, certain dust opacities and temperatures must be adopted, describing the physical properties of the observed region. The process involves calculation of the \textit{Planck} images at the \textit{Herschel} wavelengths, using the dust optical depth $\tau_{353\,\text{GHz}}$ (at 850 $\mu $m) and temperature $T$ from the \textit{Planck} observations \citep{Planck+2014} and assuming the optically-thin blackbody radiation:
\begin{equation}
I_{\nu} = \tau_{\nu_0} \left(\frac{\nu}{\nu_0}\right)^{\beta} B_{\nu}(T),
\label{equA1}
\end{equation}
where $\nu$ is the frequency corresponding to the wavebands of the \textit{Herschel} images and $B_{\nu}(T)$ is the blackbody intensity at the temperature $T$. Adopting $\beta = 2$ and $\nu_0 = 353$ GHz, we obtain the optical depth $\tau_{\nu_0}$ directly from the \textit{Planck} images and, therefore, we calculate the \textit{Planck} images at the \textit{Herschel} wavelengths. The \textit{Herschel} observation maps are then smoothed and resampled to match the \textit{Planck} pixels. The offsets between the \textit{Herschel} and \textit{Planck} images are calculated as their median differences over all pixels for each wavelength (Fig.~\ref{linfitoffsets}).
The resulting offsets of 134.2, 37.2, 12.8, and 3.5 MJy sr$^{-1}$ at 160, 250, 350, and 500 $\mu$m, respectively, were added to the \textit{Herschel} images to create the high-resolution surface density and temperature maps.

\begin{figure*}
  \centering
  \includegraphics[width=0.48\hsize]{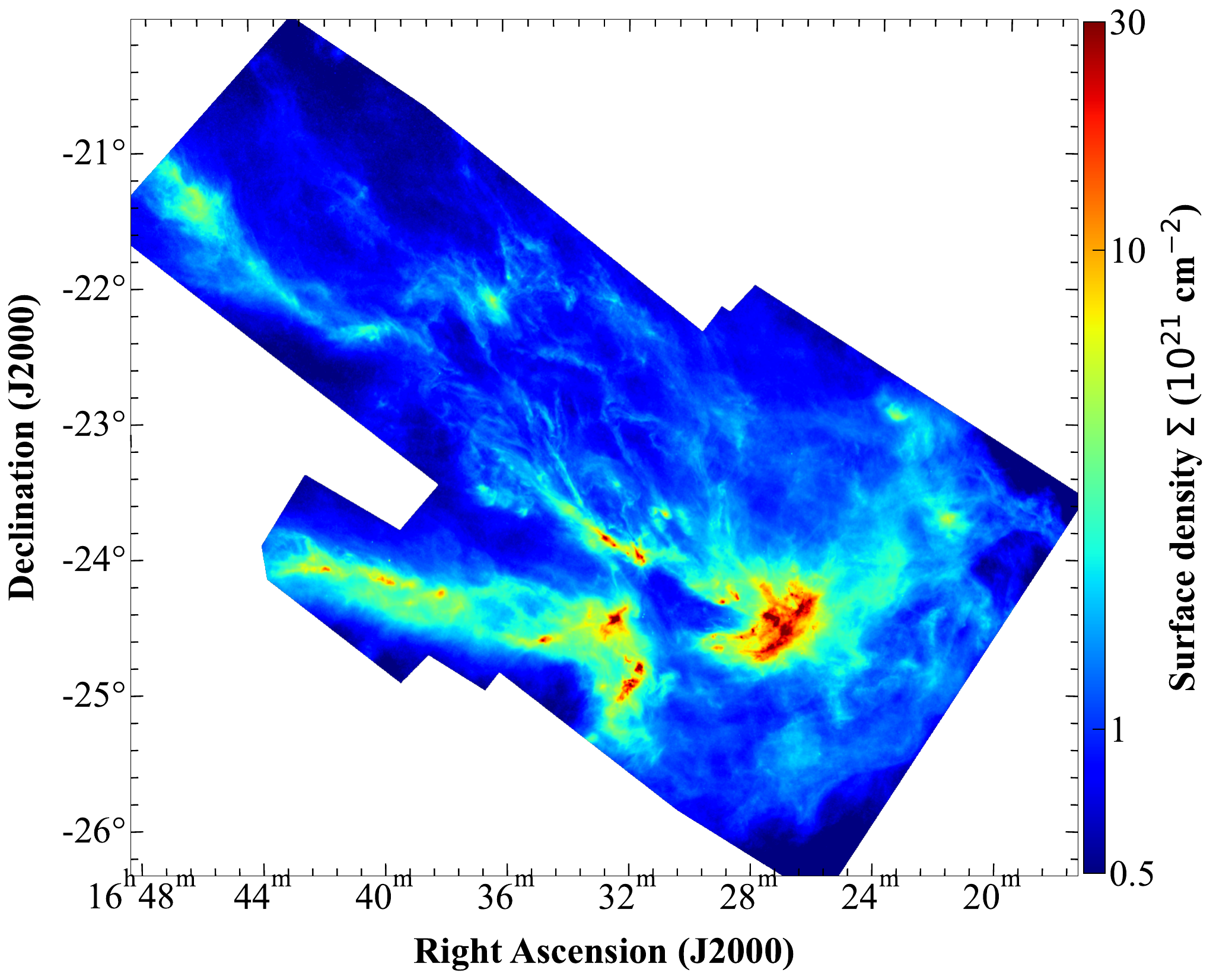}
  \includegraphics[width=0.48\hsize]{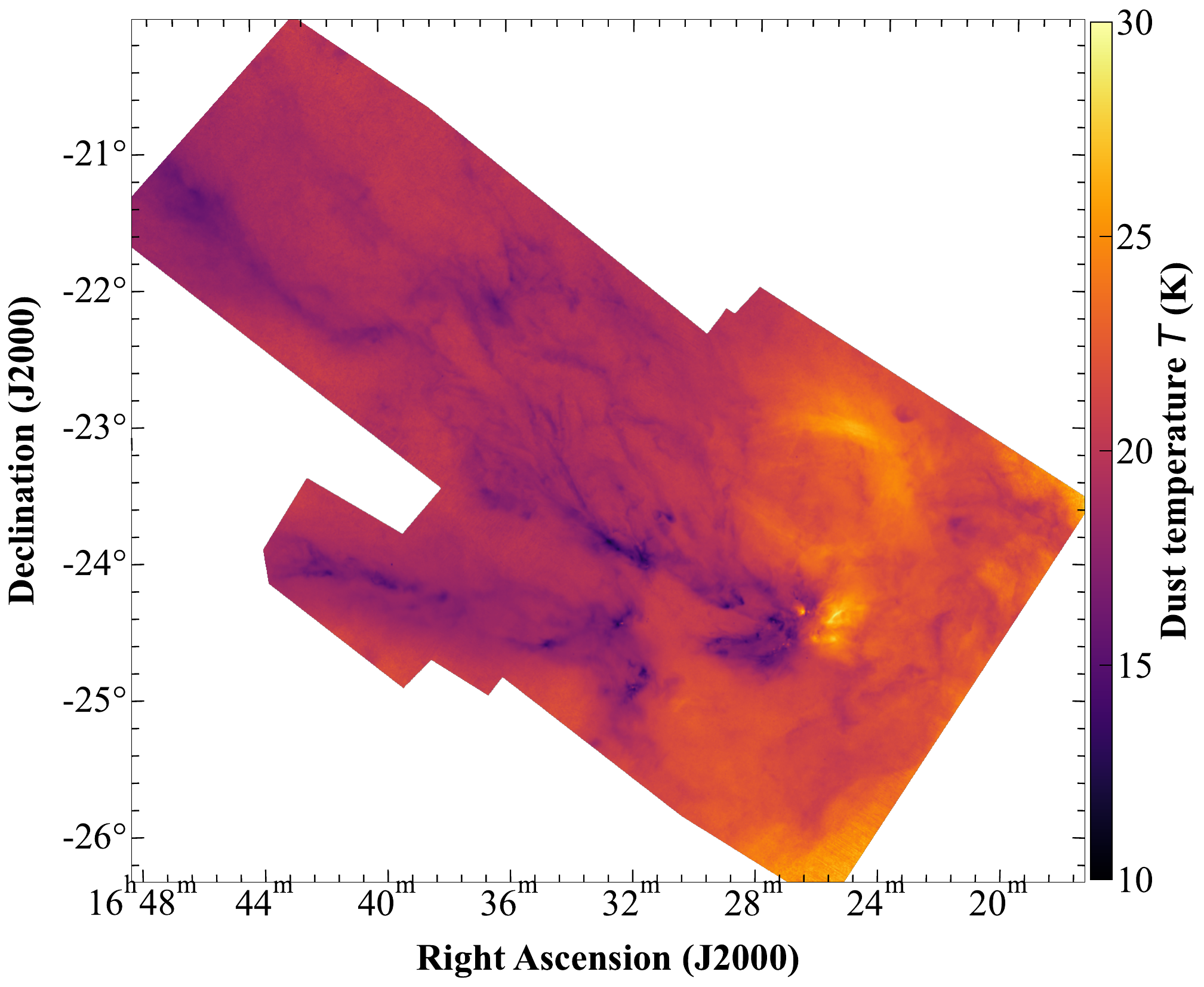}
    \caption{Maps of the high-resolution surface densities (left panel) and dust temperatures (right panel) of the Ophiuchus region with a spatial resolution of 13.5\arcsec. The map was derived using the \textit{hires} algorithm from the \textit{getsf} software \citep{Menshchikov2021method}, by processing the \textit{Herschel} multi-wavelength far-infrared images at 160, 250, 350, and 500 $\mu$m.}
  \label{fig:Surfacemap13p5}
\end{figure*}

  \begin{figure*}
       \centering
       \includegraphics[width=0.48\hsize]{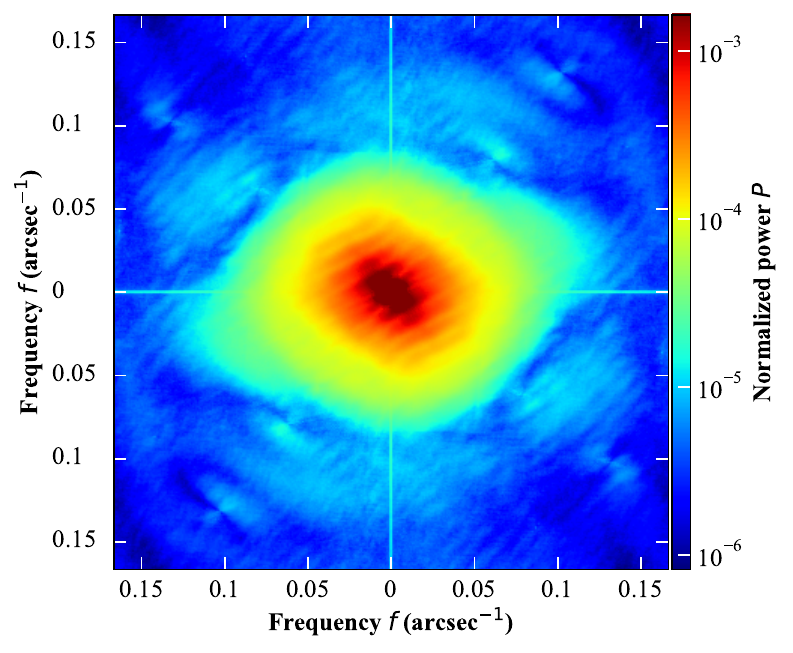}
       \includegraphics[width=0.48\hsize]{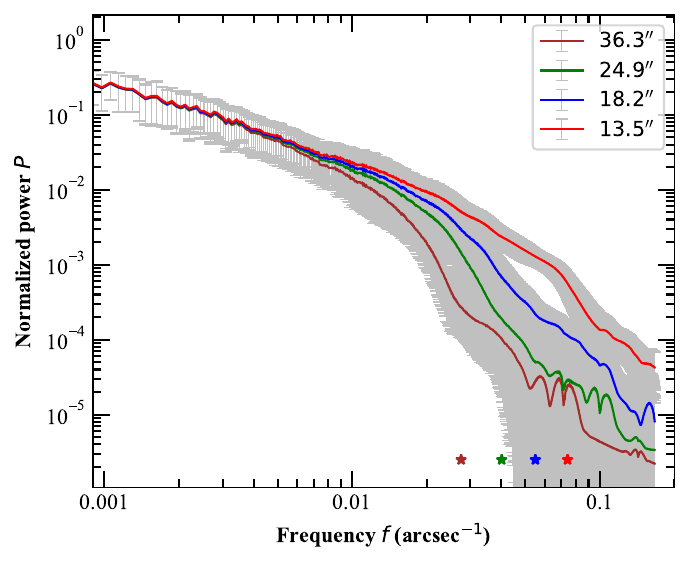}
       \caption{Fourier amplitudes (left panel) for the image of surface densities with the highest resolution of 13.5{\arcsec}. Power spectra (right panel), obtained by azimuthally-averaging the Fourier amplitudes for the surface densities with the angular resolutions of 13.5, 18.2, 24.9, and 36.3{\arcsec} (filled pentagrams), derived using the \textit{hires} method.}
       \label{fourieramp}
   \end{figure*}

The \textit{hires} algorithm derives the high-resolution surface density and temperature images from multiwavelength far-infrared observations assuming optically thin dust emission. The images are resampled to a common (the smallest) pixel size and then convolved to all available angular resolutions of the \textit{Herschel} images. The spectral shapes of pixel intensities in the observed images are then fitted with a modified blackbody using the \textit{fitfluxes} utility \citep{Men2016fitfluxes}, assuming that the dust opacity is $\kappa_{\lambda} = 0.1\,(\lambda /300\,\mu\rm{m})^{-2}$ cm$^{2}$\,g$^{-1}$ (per gram of dusty gas). The \textit{hires} algorithm obtains a series of surface density and temperature images for available combinations of the wavelengths \citep{Menshchikov2021method}. Making differential improvements to the image with the lowest angular resolution of 36.3{\arcsec}, it produces additional images with higher resolutions of 13.5, 18.2, and 24.9{\arcsec}. For an illustration, Fig.~\ref{fig:Surfacemap13p5} displays the surface density and temperature maps with a resolution of 13.5{\arcsec}.

\subsection{Consistency Checks for the Derived Surface Densities and Temperatures}

Overall compatibility of the images with added offsets can be tested with a simple approach \citep{Menshchikov2021method}. When all four images are convolved to the 36.3{\arcsec} resolution of the 500 $\mu$m image, a pixel-to-pixel SED fitting of the three pairs of images (160, 250 $\mu$m), (250, 350 $\mu$m), and (350, 500 $\mu$m), must give the same temperatures, if the images are consistent and the fitting model and assumptions are realistic. This can be verified in an average sense, using a median value of the relative differences between the derived temperature images in each pixel, for each of the pairs. For the (160, 250 $\mu$m) images, we find the median temperature of 19.55 K, for the (250, 350 $\mu$m) images, the median temperature of 19.89 K, and for the (350, 500 $\mu$m) images, the median temperature of 19.95 K. These median values imply no serious inconsistency in the images and offsets used in the derivation of the surface densities and temperatures.

   \begin{figure*}
     \centering
     \includegraphics[width=0.48\hsize]{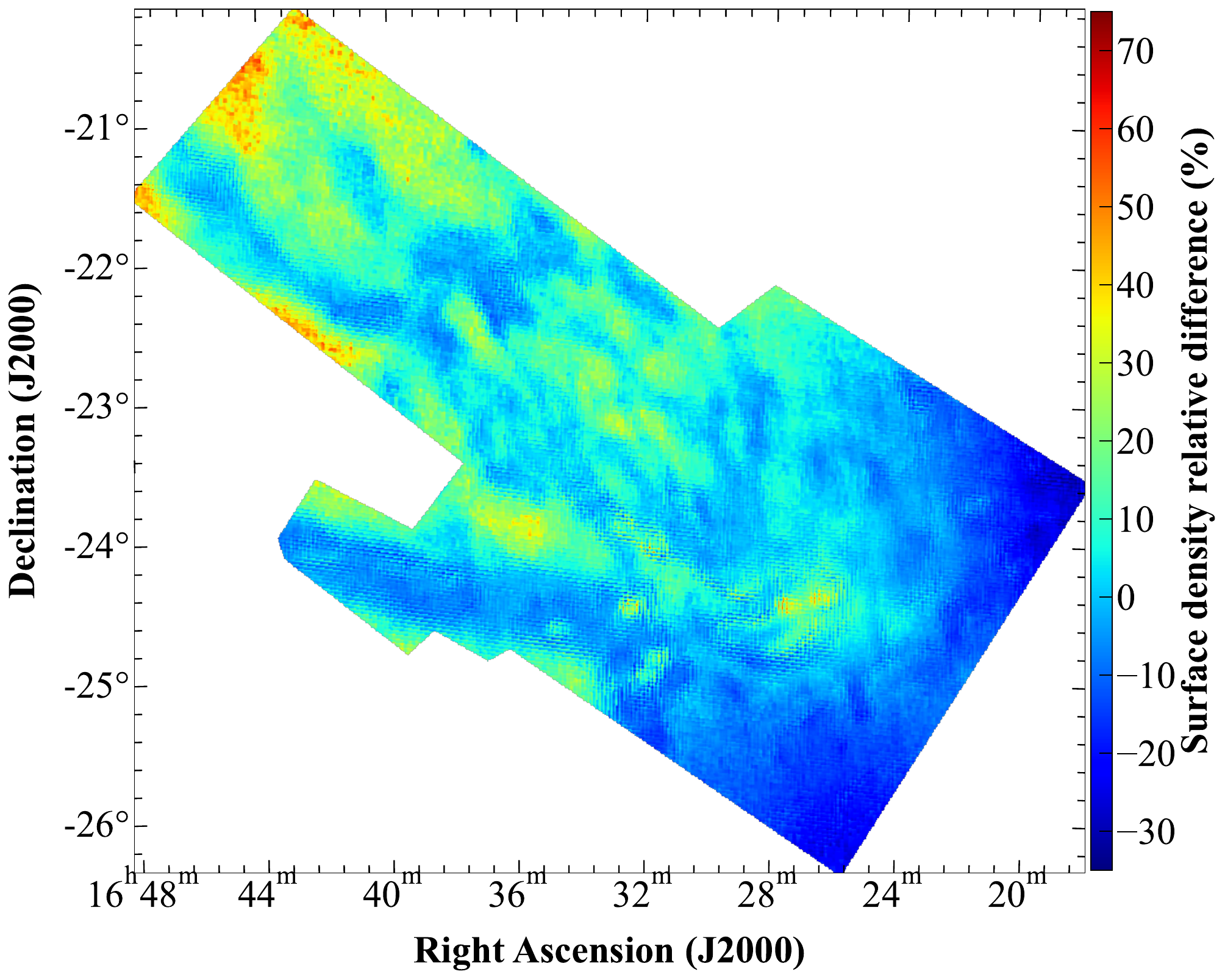}
     \includegraphics[width=0.48\hsize]{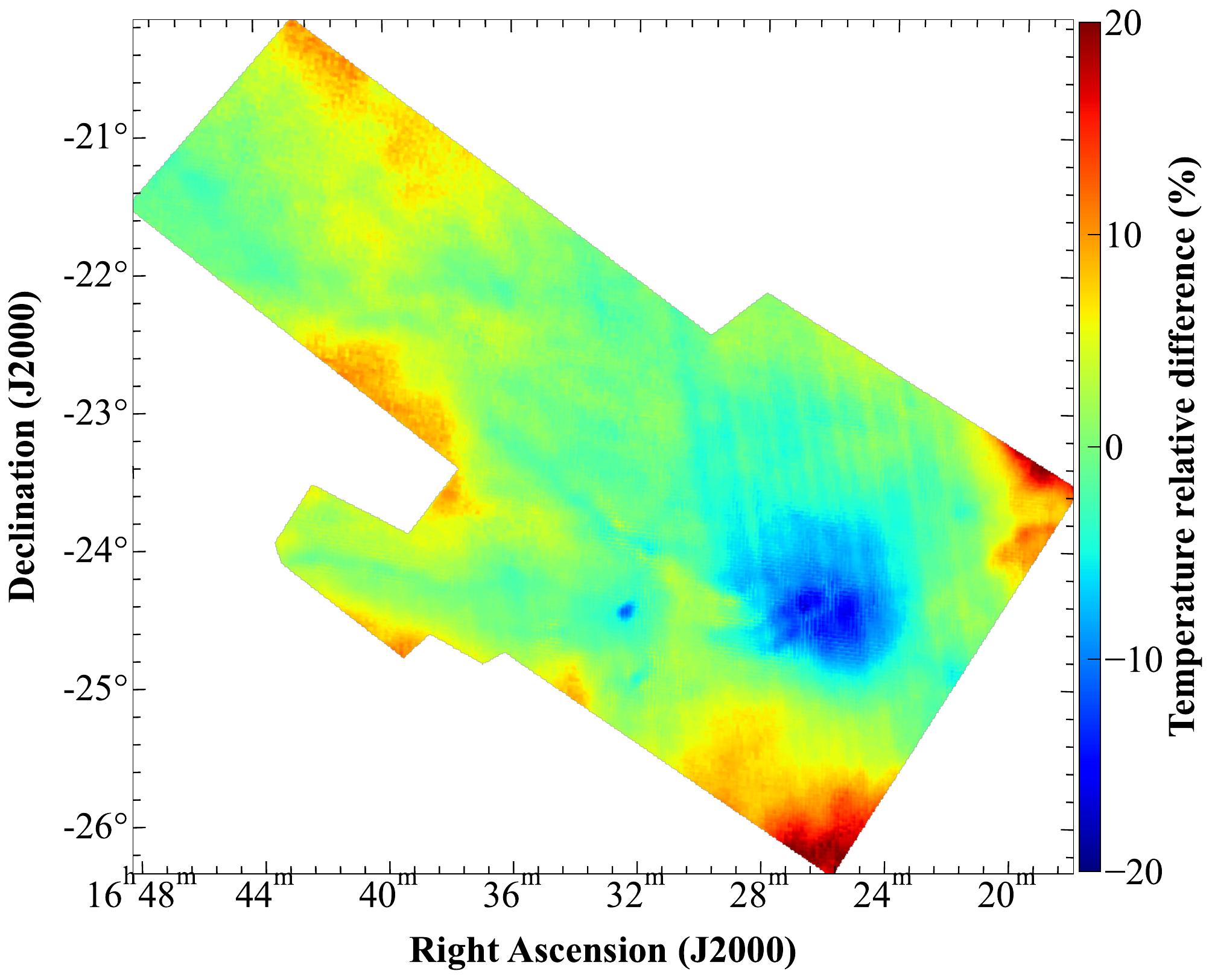}
     \caption{Image of the relative differences between the maps of surface densities (left panel) and dust temperatures (right panel) of the Ophiuchus molecular cloud, derived from the \textit{Herschel} and \textit{Planck} observations. The \textit{Herschel} map was convolved and resampled to match the \textit{Planck} angular resolution and pixel size before calculation of the relative differences.}
     \label{HPsurfaceRelative}  
   \end{figure*}

We computed Fourier amplitudes and power spectra for surface density maps of the L1688 region at resolutions of 13.5, 18.2, 24.9, and 36.3{\arcsec} (Fig.~\ref{fourieramp}) to visualize the differences between high-resolution and low-resolution images. The power spectra confirm that the higher-resolution images capture finer details with larger Fourier amplitudes at higher spatial frequencies and that at much lower spatial frequencies, representing larger-scale structures, the values are almost the same across the different resolutions. This is because the \textit{hires} method integrates the higher-resolution contributions into the lower-resolution surface densities, accumulating them to enhance the accuracy and resolution of the resulting images.

To verify consistency of the surface densities and temperatures of the Ophiuchus cloud, derived from the \textit{Herschel} observations, we computed the \textit{Planck} surface densities $\Sigma_{\rm P}$ from the 353 GHz (850 $\mu$m) image of optical depths $\tau_{353}$ \citep{Planck+2014},
\begin{equation}
    \Sigma_{\rm P} = \frac{\tau_{353}}{\kappa_{353\,} \mu_{\rm{H}_2} m_{\rm{H}}},
\end{equation}
where $\kappa_{353}$ is the dust opacity, defined in Appendix~\ref{sec:offsets_surface_densities}, at 850 $\mu$m. We smoothed the surface densities and temperatures derived from $\textit{Herschel}$ observations to the angular resolution of the $\textit{Planck}$ images and computed their relative differences in each pixel. Figure~\ref{HPsurfaceRelative} demonstrates that the differences between the two are mostly within $\sim$\,20\% in the surface densities and within $\sim$\,10\% in the dust temperatures. The largest differences are found in the low surface density areas and/or close to the edges of the \textit{Herschel} coverage (Fig.~\ref{fig:Surfacemap13p5}). However, the main area of the surface density image of the Ophiuchus molecular cloud, where we performed our source and filament extractions, has substantially smaller differences with the \textit{Planck} data.

\section{Catalogs of the Extracted Cores and Filaments in Ophiuchus}

Using the \textit{getsf} method with the \emph{Herschel} SPIRE and PACS images of the Ophiuchus molecular cloud, we extracted 882 reliable cores and 769 well-resolved filaments. A template of the online catalog with the observed properties of the cores is provided in Table~\ref{coreobs}, with one protostellar core (illustrated in Fig.~\ref{fig:protostellar}) included as an example. Tables~\ref{corederive} and \ref{filamentpara} present templates of the derived properties of cores and filaments, respectively. Tables~\ref{coreobs}, \ref{corederive}, and \ref{filamentpara} were generated by \textit{fitfluxes}, \textit{smeasure}, and \textit{fmeasure}, respectively, the utilities from the \textit{getsf} software. The notation follows the conventions from \citet{Menshchikov2021method}, and the formatting is consistent with \citet{Li+2023}. The tables illustrate the contents of the full catalog, which is available online at \url{https://www.scidb.cn/en/s/Y3AFv2}.

\onecolumn
\setcounter{table}{0}
\begin{landscape}
\tiny
\setlength{\tabcolsep}{0.8mm}{
\begin{longtable}{c|cccccccc}
\caption{\label{coreobs} Catalog of 882 Reliable Cores Identified in the Multiwavelength \emph{Herschel} Maps of the Ophiuchus Molecular Cloud (template, full catalog only provided online at \url{https://www.scidb.cn/en/s/Y3AFv2}).}\\
  \hline\noalign{\smallskip}
$n$ & RA & Dec & $f$ & $\Gamma$ & $\Xi$ & Core type & SIMBAD \\
 & (J2000, $\degr$) & (J2000, $\degr$) &  &  &  &  &  \\
(1) & (2) & (3) & (4) & (5) & (6) & (7) & (8) \\
  \hline\noalign{\smallskip}
14 & 246.7444020 & -24.7605597 & 0 & 5.788E+02 & 1.015E+04 & protostellar &
 \\
  $\cdots$ & $\cdots$ \\
  \noalign{\smallskip}\hline
\end{longtable}
\begin{longtable}{ccccccccccccccccc}
  \hline\noalign{\smallskip}
$f_{70}$ & $\Gamma_{70}$ & $\Xi_{70}$ & $F_{\textrm{P},70}$ & $\sigma_{\textrm{P},70}$ & $F_{\textrm{T},70}$ & $\sigma_{\textrm{T},70}$ & $F_{\textrm{G},70}$ & $S_{j_{\textrm{F}},70}$& $A_{70}$ & $B_{70}$ & $A_{\textrm{M},70}$ & $B_{\textrm{M},70}$ & $\omega_{\textrm{M},70}$ & $\phi_{70}$ & $A_{\textrm{F},70}$ & $B_{\textrm{F},70}$ \\
 &  &  & (Jy/beam) & (Jy/beam) & (Jy) & (Jy) & (Jy) &  & (\arcsec) & (\arcsec) & (\arcsec) & (\arcsec) & ($\degr$)  &   & (\arcsec) & (\arcsec) \\
(9) & (10) & (11) & (12) & (13) & (14) & (15) & (16) & (17) & (18) & (19) & (20) & (21) & (22) & (23) & (24) & (25) \\   
0 & 3.520E+02 & 1.005E+04 & 5.354E+00 & 1.592E-02 & 9.991E+00 & 3.027E-02 & 8.374E+00 & 1.893E+01 & 1.135E+01 & 9.725E+00 & 1.441E+01 & 1.289E+01 & 1.382E+02 & 2.553E+00 & 5.178E+01 & 4.436E+01 \\
   $\cdots$ \\
  \hline\noalign{\smallskip}
  \noalign{\smallskip}\hline
  
$f_{160}$ & $\Gamma_{160}$ & $\Xi_{160}$ & $F_{\textrm{P},160}$ & $\sigma_{\textrm{P},160}$ & $F_{\textrm{T},160}$ & $\sigma_{\textrm{T},160}$ & $F_{\textrm{G},160}$ & $S_{j_{\textrm{F}},160}$ & $A_{160}$ & $B_{160}$ & $A_{\textrm{M},160}$ & $B_{\textrm{M},160}$ & $\omega_{\textrm{M},160}$ & $\phi_{160}$ & $A_{\textrm{F},160}$ & $B_{\textrm{F},160}$ \\
 &  &  & (Jy/beam) & (Jy/beam)  & (Jy) & (Jy) & (Jy) &  & (\arcsec) & (\arcsec) & (\arcsec) & (\arcsec) & ($\degr$)  &   & (\arcsec) & (\arcsec) \\
(26) & (27) & (28) & (29) & (30) & (31) & (32) & (33) & (34) & (35) & (36) & (37) & (38) & (39) & (40) & (41) & (42) \\
  \hline\noalign{\smallskip}
0 & 2.116E+02 & 1.044E+03 & 7.451E+00 & 1.186E-01 & 9.995E+00 & 1.412E-01 & 6.514E+00 & 1.893E+01 & 1.461E+01 & 1.247E+01 & 1.764E+01 & 1.607E+01 & 1.229E+02 & 2.553E+00 & 4.942E+01 & 4.702E+01 \\  
   $\cdots$ \\
  \hline\noalign{\smallskip}
  \noalign{\smallskip}\hline

$f_{13.5{\arcsec}}$ & $\Gamma_{13.5{\arcsec}}$ & $\Xi_{13.5{\arcsec}}$ & $F_{\textrm{P},13.5{\arcsec}}$ & $\sigma_{\textrm{P},13.5{\arcsec}}$ & $F_{\textrm{T},13.5{\arcsec}}$ & $\sigma_{\textrm{T},13.5{\arcsec}}$ & $F_{\textrm{G},13.5{\arcsec}}$ & $S_{j_{\textrm{F}},13.5{\arcsec}}$ & $A_{13.5{\arcsec}}$ & $B_{13.5{\arcsec}}$ & $A_{\textrm{M},13.5{\arcsec}}$ & $B_{\textrm{M},13.5{\arcsec}}$ & $\omega_{\textrm{M},13.5{\arcsec}}$ & $\phi_{13.5{\arcsec}}$ & $A_{\textrm{F},13.5{\arcsec}}$ & $B_{\textrm{F},13.5{\arcsec}}$ \\
 &  &  & $(\textrm{cm}^{-2}$/beam) & $(\textrm{cm}^{-2}$/beam) & $(\textrm{arcsec}^{2}/\textrm{cm}^{2})$ & $(\textrm{arcsec}^{2}/\textrm{cm}^{2})$ & $(M_{\odot})$ &  & (\arcsec) & (\arcsec) & (\arcsec) & (\arcsec) & ($\degr$) &  & (\arcsec) & (\arcsec) \\
(43) & (44) & (45) & (46) & (47) & (48) & (49) & (50) & (51) & (52) & (53) & (54) & (55) & (56) & (57) & (58) & (59) \\ 
0 & 3.285E+02 & 8.382E+02 & 2.241E+24 & 8.729E+22 & 4.151E+24 & 1.439E+23 & 4.536E-02 & 4.864E+00 & 1.462E+01 & 1.371E+01 & 2.463E+01 & 2.273E+01 & 2.009E+01 & 3.528E+00 & 6.704E+01 & 6.647E+01 \\
   $\cdots$ \\
  \hline\noalign{\smallskip}
  \noalign{\smallskip}\hline
  
$f_{250}$ & $\Gamma_{250}$ & $\Xi_{250}$ & $F_{\textrm{P},250}$ & $\sigma_{\textrm{P},250}$ & $F_{\textrm{T},250}$ & $\sigma_{\textrm{T},250}$ & $F_{\textrm{G},250}$ & $S_{j_{\textrm{F}},250}$ & $A_{250}$ & $B_{250}$ & $A_{\textrm{M},250}$ & $B_{\textrm{M},250}$ & $\omega_{\textrm{M},250}$ & $\phi_{250}$ & $A_{\textrm{F},250}$ & $B_{\textrm{F},250}$ \\
 &  &  & (Jy/beam) & (Jy/beam) & (Jy) & (Jy) & (Jy) &  & (\arcsec) & (\arcsec) & (\arcsec) & (\arcsec) & ($\degr$)  &   & (\arcsec) & (\arcsec) \\
(60) & (61) & (62) & (63) & (64) & (65) & (66) & (67) & (68) & (69) & (70) & (71) & (72) & (73) & (74) & (75) & (76) \\
0 & 1.546E+02 & 3.094E+02 & 3.999E+00 & 2.117E-01 & 4.029E+00 & 1.617E-01 & 4.180E+00 & 1.893E+01 & 2.086E+01 & 1.924E+01 & 1.908E+01 & 1.780E+01 & 1.763E+02 & 2.502E+00 & 4.753E+01 & 4.715E+01 \\
   $\cdots$ \\

   \\   
  \hline\noalign{\smallskip}
  \noalign{\smallskip}\hline

$f_{350}$ & $\Gamma_{350}$ & $\Xi_{350}$ & $F_{\textrm{P},350}$ & $\sigma_{\textrm{P},350}$ & $F_{\textrm{T},350}$ & $\sigma_{\textrm{T},350}$ & $F_{\textrm{G},350}$ & $S_{j_{\textrm{F}},350}$ & $A_{350}$ & $B_{350}$ & $A_{\textrm{M},350}$ & $B_{\textrm{M},350}$ & $\omega_{\textrm{M},350}$ & $\phi_{350}$ & $A_{\textrm{F},350}$ & $B_{\textrm{F},350}$ \\
 &  &  & (Jy/beam) & (Jy/beam) & (Jy) & (Jy) & (Jy) &  & (\arcsec) & (\arcsec) & (\arcsec) & (\arcsec) & ($\degr$)  &   & (\arcsec) & (\arcsec) \\
(77) & (78) & (79) & (80) & (81) & (82) & (83) & (84) & (85) & (86) & (87) & (88) & (89) & (90) & (91) & (92) & (93) \\
0 & 9.635E+01 & 1.171E+02 & 2.521E+00 & 2.370E-01 & 2.435E+00 & 1.664E-01 & 2.524E+00 & 2.490E+01 & 2.698E+01 & 2.629E+01 & 2.406E+01 & 2.377E+01 & 6.280E+01 & 2.311E+00 & 5.762E+01 & 5.754E+01 \\
   $\cdots$ \\
  \hline\noalign{\smallskip}
  \noalign{\smallskip}\hline

$f_{500}$ & $\Gamma_{500}$ & $\Xi_{500}$ & $F_{\textrm{P},500}$ & $\sigma_{\textrm{P},500}$ & $F_{\textrm{T},500}$ & $\sigma_{\textrm{T},500}$ & $F_{\textrm{G},500}$ & $S_{j_{\textrm{F}},500}$ & $A_{500}$ & $B_{500}$ & $A_{\textrm{M},500}$ & $B_{\textrm{M},500}$ & $\omega_{\textrm{M},500}$ & $\phi_{500}$ & $A_{\textrm{F},500}$ & $B_{\textrm{F},500}$ \\
 &  &  & (Jy/beam) & (Jy/beam) & (Jy) & (Jy) & (Jy) &  & (\arcsec) & (\arcsec) & (\arcsec) & (\arcsec) & ($\degr$)  &   & (\arcsec) & (\arcsec) \\
(94) & (95) & (96) & (97) & (98) & (99) & (100) & (101) & (102) & (103) & (104) & (105) & (106) & (107) & (108) & (109) & (110) \\
0 & 4.954E+01 & 2.923E+01 & 1.300E+00 & 2.491E-01 & 1.327E+00 & 1.764E-01 &  1.395E+00 & 3.630E+01 & 4.149E+01 & 3.907E+01 & 3.621E+01 & 3.440E+01 & 2.190E+01 & 2.311E+00 & 8.402E+01 & 8.389E+01 \\
   $\cdots$ \\
  \hline\noalign{\smallskip}

\end{longtable}
}
\tablefoot{Catalog entries are as follows: 
(1) Source running number; 
(2) and (3): Centroid equatorial coordinates;
(4) Flag describing global properties over all wavelengths (global flag), 0: source is not blended with any other source in any waveband; 1: source's footprints intersect by more than 20\% in at least one waveband; 2: source's footprint area contains at least one other source; 3: source is causing a larger source to be sub-structured; 
(5) The detection significance over all wavelengths; 
(6) The monochromatic goodness (combining detection significance and signal-to-noise ratio); 
(7) Core type: protostellar, candidate or robust prestellar and unbound starless;
(8) SIMBAD infrared source counterparts within a radius of 6{\arcsec} from the centroid position of the \emph{Herschel} source; 
(9), (26), (43), (60), (77), (94): Wavelength-dependent flag; 
(10), (27), (44), (61), (78), (95): The detection significance from monochromatic single scales; 
(11), (28), (45), (62), (79), (96): The monochromatic goodness (combining detection significance and signal-to-noise ratio); 
(12), (29), (46), (63), (80), (97): The peak intensity; 
(13), (30), (47), (64), (81), (98): The error of $F_{\textrm{P},\lambda}$; 
(14), (31), (48), (65), (82), (99): The total flux; 
(15), (32), (49), (66), (83), (100): The error of $F_{\textrm{T},\lambda}$; 
(16), (33), (67), (84): The Gaussian flux; 
(50): The mass derived from the surface density image; 
(17), (34), (51), (68), (85), (102): The characteristic size of sources; 
(18)$-$(19), (35)$-$(36), (53)$-$(54), (69)$-$(70), (86)$-$(87), (103)$-$(104): The major and minor sizes at half-maximum of sources; 
(20)$-$(21), (37)$-$(38), (55)$-$(56), (71)$-$(72), (88)$-$(89), (105)$-$(106): The major and minor size from intensity moments of sources;  
(22), (39), (57), (73), (90), (107): The position angle; 
(23), (40), (58), (74), (91), (108): The footprint factor;
(24)$-$(25), (41)$-$(42), (59)$-$(60), (75)$-$(76), (92)$-$(93), (109)$-$(110): The major and minor sizes at footprint axes of sources. 
}
\end{landscape}

\setcounter{table}{1}
\begin{landscape}
\tiny
\setlength{\tabcolsep}{1.0mm}{
\begin{longtable}{c|ccccccccccccccccccc}
\caption{\label{corederive} Derived Properties of 882 Reliable Cores Identified in the Multiwavelength \emph{Herschel} Maps of Ophiuchus Molecular Cloud (template, full table only provided online at \url{https://www.scidb.cn/en/s/Y3AFv2}).}\\
  \hline\noalign{\smallskip}
  \noalign{\smallskip}\hline
  $n$ & RA & Dec & $T$ & $\sigma_{T}$ & $M$ & $\sigma_{M}$ & $L$ & $\sigma_{L}$ & $R$ & $R$ & $\omega_{\textrm{M}}$ & $A_{\textrm{F}}$ & $B_{\textrm{F}}$ & $F_{\textrm{P}}$ & $\alpha_{\rm BE}$ & Core type \\
   & (J2000, $\degr$) & (J2000, $\degr$) & (K) & (K) & $(M_\odot)$ & $(M_\odot)$ & $(L_\odot)$ & $(L_\odot)$ & (\arcsec) &  (pc) & ($\degr$) & ($\arcsec$) & ($\arcsec$) & ($\textrm{cm}^{-2}$) &   \\
  (1) & (2) & (3) & (4) & (5) & (6) & (7) & (8) & (9) & (10) & (11) & (12) & (13) & (14) & (15) & (16) & (17) \\
    \hline\noalign{\smallskip}
  14 & 246.7444020 & -24.7605597 & 1.867E+01 & 1.680E+00 & 2.710E-02 & 7.830E-03 & 1.106E-01 & 1.212E-02 & 1.416E+01 & 9.611E-03 & 2.009E+01 & 6.704E+01 & 6.647E+01 & 1.085E+22 & 3.630E+00 & protostellar \\
  25 & 246.7431627 & -24.5717742 & 1.077E+01 & 6.602E-01 & 2.874E+00 & 7.315E-01 & 4.293E-01 & 3.364E-02 & 6.795E+01 & 4.612E-02 & 1.021E+02 & 1.190E+02 & 1.184E+02 & 3.507E+22 & 3.200E-01 & prestellar  \\
  52 & 248.2805876 & -22.8453093 & 3.000E+01 & 0.000E+00 & 5.392E-03 & 7.370E-05 & 1.193E-01 & 1.400E-02 & 1.350E+01 & 9.163E-03 & 1.293E+02 & 3.419E+01 & 3.396E+01 & 2.177E+21 & 2.850E+00 & unbound  \\
  $\cdots$ & $\cdots$ \\
  \hline\noalign{\smallskip}
  Max    & $-$ & $-$ & 3.636E+02 & $-$ & 1.287E+01 & $-$ & 4.066E+01 & $-$ & 1.720E+02 & 1.167E-01 & 1.799E+02 & 2.515E+02 & 2.508E+02 & 1.706E+26 & 3.044E+03 & $-$ \\
  Min    & $-$ & $-$ & 6.000E+00 & $-$ & 3.083E-05 & $-$ & 3.865E-04 & $-$ & 1.350E+01 & 9.163E-03 & 6.100E-01 & 3.120E+01 & 3.120E+01 & 1.451E+22 & 2.000E-02 & $-$ \\
  Median & $-$ & $-$ & 1.585E+01 & $-$ & 1.625E-02 & $-$ & 2.978E-02 & $-$ & 3.879E+01 & 2.633E-02 & 9.090E+01 & 7.211E+01 & 7.116E+01 & 1.058E+23 & 3.259E+01 & $-$ \\  
  Mean   & $-$ & $-$ & 1.746E+01 & $-$ & 2.072E-01 & $-$ & 2.222E-01 & $-$ & 4.295E+01 & 2.915E-02 & 9.105E+01 & 8.020E+01 & 7.985E+01 & 9.199E+23 & 6.530E+01 & $-$ \\
  \hline
\end{longtable}
}
\tablefoot{Catalog entries are as follows: 
(1): Source running number; 
(2): and (3): Centroid equatorial coordinates;
(4): Derived dust temperature from SED fitting; 
(5): Uncertainty of derived temperature; 
(6): Derived total mass (gas and dust) from SED fitting; 
(7): Uncertainty of derived total mass; 
(8): Derived bolometric luminosity from SED fitting; 
(9): Uncertainty of derived luminosity;
(10), (11): Geometrical mean of the major and minor FWHM sizes, expressed in both parsecs (pc) and arcseconds (arcsec).
(12): Position angle;
(13): Full major axis of an elliptical footprint;
(14): Full minor axis of an elliptical footprint;
(15): Peak surface density;
(16): Bonnor-Ebert stability parameter;
(17): Core type: protostellar, candidate or robust prestellar and unbound starless.
The bottom four rows of the table present the maximum, minimum, median, and mean values for all 882 reliable cores.
}
\tiny
\setlength{\tabcolsep}{1.0mm}{
\begin{longtable}{c|ccccccccccccccccc}
\caption{\label{filamentpara} Catalog of the 769 Well-Resolved Filaments Identified in the Multiwavelength \emph{Herschel} Maps of the Ophiuchus Molecular Cloud (template, full catalog only provided online at \url{https://www.scidb.cn/en/s/Y3AFv2}).}\\
  \hline\noalign{\smallskip}
  \noalign{\smallskip}\hline
  $n$ & RA & Dec & $W$ & $L$ & $W$ & $W_{\alpha}$ & $W_{\beta}$ & $\overline{W}$ & $\overline{W}_{\alpha}$ & $\overline{W}_{\beta}$ & $\varsigma _{\overline{W}}$  & $\varsigma _{\overline{W}_{\alpha}}$ & $\varsigma _{\overline{W}_{\beta}}$ & $N_{\rm W}$ & $\Omega_{\overline{W}}$ & $\Omega_{\overline{W}_{\alpha}}$ & $\Omega_{\overline{W}_{\beta}}$ \\
   & (J2000, $\degr$) & (J2000, $\degr$) & (\arcsec) & (pc) & (pc) & (pc) & (pc) & (pc) & (pc) & (pc) & (pc) & (pc) & (pc) &  &  \\
  (1) & (2) & (3) & (4) & (5) & (6) & (7) & (8) & (9) & (10) & (11) & (12) & (13) & (14) & (15) & (16) & (17) & (18) \\
    \hline\noalign{\smallskip}
  1 & 250.6506958 &  -24.2532806 & 2.552E+02 & 1.126E-01 & 1.782E-01 & 1.887E-01 & 1.683E-01 & 4.667E-01 & 2.125E-01 & 1.025E+00 & 2.997E-01 & 7.332E-02 & 1.518E+00 &  76  & 1.557E+00 & 2.898E+00 & 6.753E-01 \\
  2 & 250.7283042 &  -24.1022000 & 2.354E+02 & 4.359E-01 & 1.644E-01 & 9.162E-02 & 2.949E-01 & 1.890E-01 & 1.061E-01 & 3.365E-01 & 9.002E-02 & 3.487E-02 & 8.321E-02 & 293  & 2.099E+00 & 3.043E+00 & 4.044E+00 \\
  3 & 250.8008458 &  -24.0477167 & 4.435E+02 & 1.481E-01 & 3.096E-01 & 5.870E-01 & 1.633E-01 & 3.399E-01 & 6.012E-01 & 1.922E-01 & 1.650E-01 & 7.576E-02 & 7.200E-02 &  99  & 2.060E+00 & 7.935E+00 & 2.670E+00 \\
  4 & 250.6182250 &  -24.1008389 & 5.580E+02 & 1.136E-01 & 3.895E-01 & 6.496E-01 & 2.336E-01 & 4.828E-01 & 6.073E-01 & 3.839E-01 & 3.171E-01 & 4.448E-01 & 2.728E-01 &  77  & 1.522E+00 & 1.366E+00 & 1.407E+00 \\    
  $\cdots$ & $\cdots$ \\
  \hline\noalign{\smallskip}
 \end{longtable}
 
 \begin{longtable}{cccccccccccccccccc}  
  \hline\noalign{\smallskip}
  $\Sigma_{\textrm{H}_{2}}$ & $\varsigma _{\Sigma_{\textrm{H}_{2}}}$ & $\Lambda^{\textrm{P}}$ & $\Lambda^{\textrm{P}}_{\alpha}$ & $\Lambda^{\textrm{P}}_{\beta}$ & $\Lambda^{\textrm{M}}$ & $\Lambda^{\textrm{M}}_{\alpha}$ & $\Lambda^{\textrm{M}}_{\beta}$ & $M$ & $M_{\alpha}$ & $M_{\beta}$ & $\varsigma _{\Lambda^{\textrm{P}}}$ & $\Omega_{\Lambda^{\textrm{P}}}$ & $T$ & $\varsigma _{T}$ & $C$ & $\varsigma _{C}$ \\
   $(\textrm{cm}^{-2})$ & $(\textrm{cm}^{-2})$ & $(M_{\odot}/\textrm{pc})$ & $(M_{\odot}/\textrm{pc})$ & $(M_{\odot}/\textrm{pc})$ & $(M_{\odot}/\textrm{pc})$ & $(M_{\odot}/\textrm{pc})$ & $(M_{\odot}/\textrm{pc})$ & $(M_{\odot})$ & $(M_{\odot})$ & $(M_{\odot})$ & $(M_{\odot}/\textrm{pc})$ &  & (K) & (K) &  & \\
  (19) & (20) & (21) & (22) & (23) & (24) & (25) & (26) & (27) & (28) & (29) & (30) & (31) & (32) & (33) & (34) & (35) \\
  \hline\noalign{\smallskip}
  3.082E+20 & 1.711E+19 & 1.149E+00 & 1.206E+00 & 1.080E+00 & 1.653E+00 & 2.148E+00 & 1.159E+00 & 1.861E-01 & 2.417E-01 & 1.305E-01 & 1.138E-01 & 1.009E+01 & 1.900E+01 & 1.061E-01 & 2.314E-01 & 1.260E-02 \\
  3.200E+21 & 5.168E+20 & 1.420E+01 & 1.105E+01 & 1.774E+01 & 1.840E+01 & 1.504E+01 & 2.176E+01 & 8.022E+00 & 6.555E+00 & 9.488E+00 & 3.051E+00 & 4.656E+00 & 1.728E+01 & 2.512E-01 & 2.526E+00 & 2.665E-01 \\
  1.763E+21 & 1.665E+20 & 7.486E+00 & 6.854E+00 & 7.940E+00 & 1.054E+01 & 8.688E+00 & 1.239E+01 & 1.561E+00 & 1.287E+00 & 1.835E+00 & 6.272E-01 & 1.194E+01 & 1.770E+01 & 9.776E-02 & 1.634E+00 & 1.003E-01 \\
  4.284E+21 & 5.463E+20 & 1.973E+01 & 1.724E+01 & 1.741E+01 & 3.046E+01 & 3.978E+01 & 2.113E+01 & 3.461E+00 & 4.520E+00 & 2.401E+00 & 5.117E+00 & 3.857E+00 & 1.637E+01 & 3.589E-01 & 3.046E+00 & 3.898E-01 \\
  $\cdots$ \\
  \hline\noalign{\smallskip}
\end{longtable}
}
\tablefoot{Filaments are measured on 13.5\arcsec\, resolution surface density map. Catalog entries are as follows: 
(1) Filament number; 
(2) and (3): Equatorial coordinates of the midpoint; 
(4) and (6): Median FWHM measured from both side in radial direction (different units); 
(5): Filament length along the skeleton ; 
(7) and (8): Median FWHM measured separately from the left and right sides in the radial direction; 
(9): Mean FWHM measured from both side in radial direction; 
(10) and (11): Mean FWHM measured from the left and right side in radial direction; 
(12), (13), (14): Standard deviation about $\overline{W}$, $\overline{W}_{\alpha}$ and $\overline{W}_{\beta}$; 
(15) Number of points used for $\overline{W}$ and $\varsigma _{\overline{W}}$; 
(16) Signal-to-noise ratio of $\overline{W}$/$\varsigma _{\overline{W}}$; 
(17) and (18): Signal-to-noise ratio of $\overline{W}_{\alpha}$/$\varsigma _{\overline{W}_{\alpha}}$ and $\overline{W}_{\beta}$/$\varsigma _{\overline{W}_{\beta}}$; 
(19) Surface density along the filament skeleton; 
(20) Standard deviation about $\Sigma_{\textrm{H}_{2}}$; 
(21) Linear density measured from median integrated radial profile; 
(22) and (23): Linear density measured separately from the left and right median integrated radial profiles; 
(24) Linear density derived from $M/L$; 
(25) and (26): Linear density derived from from $M_{\alpha}/L$ and $M_{\beta}/L$; 
(27) Mass derived from both sides; 
(28) and (29): Mass derived separately from the left and right side; 
(30) Standard deviation about $\Lambda^{P}$; 
(31) Signal-to-noise ratio of $\Lambda^{P}$/$\varsigma _{\Lambda^{P}}$; 
(32) Dust temperature along the skeleton;
(33) Standard deviation about $T$; 
(34) Contrast along the skeleton.
(35) Standard deviation about $C$.
}

\end{landscape}
\end{appendix}
\end{document}